\documentclass[10pt]{article}
\usepackage{a4wide}
\usepackage{color}
\usepackage{cite}
\usepackage{graphicx}

\usepackage{amssymb}
\usepackage{amsmath}
\usepackage{epstopdf}
\usepackage{soul}

\newcommand{\dd}{{\rm{d}}} 
\newcommand{\rovno}{\!\!\!\!& = &\!\!\!\!}
\newcommand{\F}{F} 

\begin{document}

\title{
All solutions of Einstein's equations in 2+1 dimensions: $\Lambda$-vacuum, pure radiation, or gyratons
}

\author{
Ji\v{r}\'{\i} Podolsk\'y${}^a$,
Robert \v{S}varc${}^a$ and
Hideki Maeda${}^b$
\thanks{{\tt
podolsky@mbox.troja.mff.cuni.cz,
robert.svarc@mff.cuni.cz,
h-maeda@hgu.jp
}}
\\ \ \\ \ \\
${}^a$ Institute of Theoretical Physics, Charles University, \\
V~Hole\v{s}ovi\v{c}k\'ach 2, 18000 Prague 8, Czech Republic.
\\ \ \\
${}^b$ Department of Electronics and Information Engineering, \\
Hokkai-Gakuen University, Sapporo 062-8605, Japan.
}

\maketitle

\begin{abstract}
Under a weak assumption of the existence of a geodesic null congruence, we present the \emph{general} solution of the Einstein field equations in three dimensions with any value of the cosmological constant, admitting an aligned null matter field, and also gyratons (a matter field in the form of a null dust with an additional internal spin). The general local solution consists of the expanding Robinson--Trautman class and the non-expanding Kundt class. The gyratonic solutions reduce to spacetimes with a pure radiation matter field when the spin is set to zero. Without matter fields, we obtain new forms of the maximally symmetric vacuum solutions. We discuss these complete classes of solutions and their various subclasses. In particular, we identify the gravitational field of an arbitrarily accelerating source (the Kinnersley photon rocket, which reduces to a Vaidya-type non-moving object) in the Robinson--Trautman class, and pp-waves, vanishing scalar invariants (VSI) spacetimes, and constant scalar invariants (CSI) spacetimes in the Kundt class.
\end{abstract}

\vfil\noindent
PACS class:  04.20.Jb, 04.50.--h, 04.40.Nr


\bigskip\noindent
Keywords: general 3D geometries, Robinson--Trautman class, Kundt class, exact solutions of Einstein's equations, cosmological constant, null matter, gyratons
\vfil
\eject

\tableofcontents

\newpage

\section{Introduction}
Classical general relativity has passed all the experimental and observational tests so far.
Nevertheless, its inadequateness to \emph{fully} describe our universe has been proven by the celebrated singularity theorems already in the 1970s~\cite{penrose1969,penrose1979} and recently generalized to low-regularity geometries
\cite{KunzingerSteinbauerStojkovicVickers:2015,KunzingerSteinbauerVickers:2015,GrafGrantKunzingerSteinbauer:2017}.
According to them, curvature singularities appearing in many exact solutions, such as the Schwarzschild black-hole or Friedmann--Lema\^{\i}tre--Robertson--Walker cosmological models, are not specific consequences of their highly symmetric nature, but are quite natural in generic solutions. In such ultra high-curvature regions near the Big Bang or deep inside black holes, the classical Einstein's field equations are no more valid and quantum effects should dominate.
Therefore, quantum theory of gravity is required to describe the whole universe in a seamless manner, which is one of the most important unsolved problems in modern physics.

Such a theory of quantum gravity governs the wave function of the spacetime which encodes all the possible topological configurations of the classical solutions. Classically, the Einstein equations determine the spacetime locally, while the topological degrees of freedom are reflected in the domains of coordinates. Indeed, one can construct globally different geometries from locally identical spacetimes by different ways of identification. In four and higher dimensions, the number of independent components of the Riemann tensor is larger than those of the Ricci tensor~\cite{wald}, which allows different solutions in vacuum distinct from  Minkowski (such as the famous Schwarzschild or Kerr black hole). One of the difficulties of quantum gravity is the fact that both \emph{local dynamical} degrees of freedom and \emph{global topological} degrees of freedom come into play simultaneously.

In this regard, 3-dimensional spacetimes are drastically different because the numbers of the independent components of the Riemann tensor and the Ricci tensor are equal.
Hence, the Ricci tensor governed by the Einstein equations completely determines the spacetime curvature locally. In other words, there is \emph{no dynamical degree of freedom in three spacetime dimensions}. Consequently, the general vacuum solution of Einstein's equations is the maximally symmetric spacetime, namely Minkowski, de~Sitter (dS), or anti-de~Sitter (AdS) for zero, positive, or negative cosmological constant $\Lambda$, respectively.

For this reason of local triviality, 3-dimensional general relativity has attracted much attention as a classroom to study how to handle the global and topological degrees of freedom in the context of quantum gravity~\cite{carlip}. Many explicit exact classical solutions have been recently summarized and classified in a comprehensive monograph \cite{GarciaDiaz:2017}. In particular, in the presence of a negative $\Lambda$, the classical theory admits a black-hole configuration with horizons given by nontrivial identifications, even though the spacetime is locally AdS~\cite{btz1,btz2}.
This so-called BTZ black hole shares similar properties to the standard Kerr--AdS black hole in four dimensions, and has been intensively investigated to reveal the quantum aspects of black holes. The BTZ black hole has been generalized to the presence of the Maxwell field~\cite{chargedbtz} and with a null dust fluid (pure radiation)~\cite{husain1994}, which represent a charged black hole and formation of a black hole, respectively.
However, it is still not clear if these are all possible solutions in such non-vacuum systems.
To obtain more general classical solutions may bring us many insights regarding quantum gravity in the presence of matter. The topic of the present paper is along this direction of research.
Indeed, we present here \emph{the most general solution of the Einstein equations with a cosmological constant $\Lambda$ and some kind of matter fields in three dimensions, under only a few weak assumptions}.

Specifically, in section~\ref{sec_geom}, we prove that (virtually) all ${D=3}$ Lorentzian geometries belong either to the family of Robinson--Trautman spacetimes or to the family of Kundt spacetimes. Appendix~A contains the Christoffel symbols and all components of the Riemann and Ricci tensors for such a general metric. In appendices~B and~C, we systematically integrate the Einstein field equations for both the disjunct Robinson--Trautman and Kundt families, respectively. The matter content we consider in section~\ref{Einstein} is the mixture of an aligned pure radiation\footnote{The general solution with a timelike dust fluid was obtained in~\cite{BarrowShawTsagas:2006}.} and a gyratonic field (null matter with internal spin/helicity).  The most general solutions are summarized and described in sections \ref{sec:discussionRT} and \ref{sec:discussionKundt}, respectively. These are 3-dimensional counterparts of the classic Robinson--Trautman solutions \cite{RobTra60,RobTra62} and the Kundt solutions \cite{Kundt:1961} in four dimensions. In ${D=4}$, they form two huge classes of exact solutions containing, as special cases, many spacetimes, in particular various highly symmetric backgrounds, black holes, accelerated sources, exact gravitational waves, and their combinations. (See \cite{Stephani:2003,GriffithsPodolsky:2009} for a review.)

Interestingly, in higher dimensions ${D>4}$, the Robinson--Trautman solutions are more constrained \cite{PodOrt06,OrtPodZof08} so that they basically reduce to highly symmetric and well-known solutions such as the generalizations of the Schwarzschild, Reissner--Nordstr\"om, and Vaidya solutions (however, for $p$-form fields, there are some interesting nontrivial solutions \cite{OrtaggioPodolskyZofka:2015}). The large complementary Kundt class of solutions in higher dimensions was presented in \cite{PodolskyZofka:2009}. As we will show here, the Robinson--Trautman solutions in ${D=3}$ are {\it less} restrictive, in contrast to ${D \ge 4}$. And, surprisingly, we can write down the general solution of the Einstein equations in a closed form starting from only weak assumptions.

\section{All geometries in 2+1 dimensions}

Let us consider a general 3-dimensional Lorentzian spacetime $({\cal M},g)$ with the metric signature ${(-++)}$. For the curvature tensors, we shall adopt the convention ${[\nabla _a ,\nabla_b]\,V^c ={R^c}_{dab}V^d}$ and ${R_{ab}={R^c}_{acb}}$, and we choose the units such that ${c=1=G}$.

\subsection{Uniqueness of Robinson--Trautman and Kundt metrics}
\label{sec_geom}

Consider an arbitrary null vector field $\mathbf{k}$. At any point we can set up a triad ${\mathbf{e}_{I}\equiv\left\{\mathbf{k},\,\mathbf{l},\,\mathbf{m}\right\}}$ of two null vectors, namely ${\mathbf{k}\cdot\mathbf{k}=0=\mathbf{l}\cdot\mathbf{l}}$, and one spatial vector $\mathbf{m}$ orthogonal to both $\mathbf{k}$ and $\mathbf{l}$ (${\mathbf{k}\cdot\mathbf{m}=0=\mathbf{l}\cdot\mathbf{m}}$), normalized as
\begin{equation}
\mathbf{k}\cdot\mathbf{l}=-1 \,, \qquad \mathbf{m}\cdot\mathbf{m}=1  \,.
\label{triad}
\end{equation}
With respect to this triad, the covariant derivative of the vector field $\mathbf{k}$ can be decomposed as~\cite{OrtaggioPravdaPravdova:2013}
\begin{equation}
k_{a;b}=K_{ll}\,k_{a}k_{b}+K_{lk}\,k_{a}l_{b}-K_{lm}\,k_{a}m_{b}
-K_{ml}\,m_{a}k_{b}-K_{mk}\,m_{a}l_{b}+K_{mm}\,m_{a}m_{b} \,, \label{CovDerDecomp}
\end{equation}
where the scalar coefficients $K_{I\!J}$ are defined by ${K_{I\!J}\equiv k_{a;b}\,e^a_{I}e^b_{\!J}}$. Notice that the terms in (\ref{CovDerDecomp}) with ${K_{kk}}$, ${K_{kl}}$, and ${K_{km}}$ are zero due to the constant (null) norm of the vector ${\mathbf{k}}$ implying ${k_{a;b}\,k^a=0}$.

To characterize properties of the null congruence generated by $\mathbf{k}$ we employ the 2+1 analogy of classic \emph{optical scalars} (twist, shear, expansion) encoded in the \emph{optical matrix} $\rho$ defined as
\begin{equation}
\rho \equiv K_{mm}= k_{a;b}\,m^a m^b \,.
\end{equation}
(See \cite{OrtaggioPravdaPravdova:2013} for more details.) Recall that these optical scalars have an invariant meaning (in the sense that they are independent of the choice of the vectors $\mathbf{l}$ and $\mathbf{m}$ of the triad ${\mathbf{e}_{I}}$) only if the null vector field $\mathbf{k}$ is \emph{geodesic}, that is ${k_{a;b}\,k^b\propto k_a}$. In such a case, the decomposition (\ref{CovDerDecomp}) further simplifies since
\begin{equation}
K_{mk}\equiv k_{a;b}\,m^ak^b=0 \,.
\label{geodeticity}
\end{equation}

Now, as a \emph{unique property of spacetimes in} ${D=3}$, the optical matrix ${\rho}$ is simply of rank ${1\times 1}$, i.e., it \emph{degenerates to a single scalar function}. Consequently, both its antisymmetric part (twist) and symmetric traceless part (shear) trivially vanish. \emph{Any} 2+1 geometry is thus necessarily \emph{twist-free} and \emph{shear-free}, with respect to the geodesic field $\mathbf{k}$. The only possibly non-trivial optical scalar is \emph{expansion} ${\Theta\equiv\rho}$. By these arguments, we have proved the following uniqueness theorem:\footnote{This observation was already made in 2010 by Chow, Pope and Sezgin~\cite{ChowPopeSezgin:2009}.}

\vspace{5mm}
\noindent \textbf{Theorem~1:} The only possible 3-dimensional Lorentzian geometries admitting a geodesic null vector field are either \emph{expanding} geometries of the \emph{Robinson--Trautman type} (with ${\Theta\not=0}$) or \emph{non-expanding} geometries of the \emph{Kundt type} (with ${\Theta=0}$). They are necessarily twist-free and shear-free.
\vspace{5mm}

Interestingly, for 2+1 geometries, the single assumption on \emph{geodesicity}\footnote{Even this weak assumption seems to be superficial. See Remark at the end of this section.} of the null congruence generated by the field~$\mathbf{k}$  is a necessary and \emph{sufficient} condition for such a congruence  to be (locally) \emph{orthogonal to a family of null hypersurfaces} ${u=\hbox{const.}}$ To prove this statement, we employ the Frobenius's theorem in the form
\begin{equation}
k_{[a;b}k_{c]}=0 \qquad \Leftrightarrow \qquad k_a \propto u_{,a}\,. \label{FrobeniusTheorem}
\end{equation}
(See Eq.~(2.45) in \cite{Stephani:2003}.)
In terms of the decomposition (\ref{CovDerDecomp}), the left-hand side of the
theorem (\ref{FrobeniusTheorem}) reduces to  ${K_{mk}\,m_{[a}l_{b}k_{c]}=0}$. This implies ${K_{mk}=0}$, that is geodesicity because (\ref{CovDerDecomp}) gives
\begin{equation}
k_{a;b}\,k^b=-K_{lk}\,k_a+K_{mk}\,m_a=-K_{lk}\,k_a
\end{equation}
due to (\ref{geodeticity}). To summarize:

\vspace{5mm}
\noindent \textbf{Theorem~2:} A null vector field in any 3-dimensional Lorentzian geometry is \emph{geodesic} if, and only if, it is \emph{hypersurface-orthogonal}.
\vspace{5mm}

Let us emphasize that this statement is \emph{not} true in  dimensions ${D>3}$ because antisymmetric part of the optical matrix (twist) is then also involved~\cite{Stephani:2003,OrtaggioPravdaPravdova:2013}.

\vspace{5mm}
\noindent \textbf{Remark:} A geodesic null vector field~$\mathbf{k}$ \emph{can (locally) be constructed in any ``reasonable'' ${D=3}$ Lorentzian geometry}. Indeed, it is possible to take a  2-dimensional spatial surface $\Sigma$, at fixed initial time, and an arbitrary null vector at its single point ${P\in\Sigma}$. This vector can be Lie-dragged along arbitrary spatial coordinates of the surface, forming thus a vector field on $\Sigma$. Taking all these null vectors on $\Sigma$ as ``initial data'', one generates a congruence of null geodesics (due to standard existence theorem for ordinary differential equations yielding geodesics). Their tangent vectors everywhere in the spacetime domain form a geodesic null vector field.
\vspace{5mm}

\subsection{Canonical coordinates}
These two theorems combined together allow us to construct convenient \emph{canonical coordinates} for all ${D=3}$ spacetimes admitting a geodesic null vector field  $\mathbf{k}$. Due to Theorem~2, these spacetimes are always foliated by a family of null hypersurfaces ${u=\hbox{const.}}$ (such that ${k_a \propto u_{,a}}$), and we naturally choose the coordinate $u$ to label them smoothly. Without loss of generality we may also assume that the congruence of null geodesics generated by $\mathbf{k}$ is affinely parametrized by a parameter~$r$, and ${k_a = -u_{,a}}$. We choose the affine parameter $r$ along the congruence as the second coordinate on the manifold. The remaining spatial coordinate, denoted as~$x$, then span the 1-dimensional ``transverse'' subspace with constant $u$ and $r$. On any null hypersurface given by fixed $u$, the coordinate $x$ is chosen to be constant along each null geodesic generated by $\mathbf{k}$, i.e., ${\mathbf{k}=\mathbf{\partial}_r}$ with ${k^a=g^{ab}k_b= -g^{ab}u_{,b}= -g^{au}}$. In such canonical coordinates $(r,\,u,\,x)$, the specific contravariant metric components $g^{ab}$ thus must have the form ${g^{uu}=0}$, ${g^{xu}=0}$, and ${g^{ru}=-1}$. The only non-vanishing contravariant metric components are $g^{xx}$, ${g^{rx}}$, ${g^{rr}}$,  and ${g^{ru}=-1}$. By inverting the matrix $g^{ab}$ we obtain the following nontrivial metric components:
\begin{equation}
g_{xx}=1/g^{xx}\,, \qquad
g_{ur}=-1\,, \qquad
g_{ux}= g_{xx}\,g^{rx} \,, \qquad
g_{uu}= -g^{rr}+g_{xx}\,{(g^{rx})}^2 \,. \label{CovariantMetricComp}
\end{equation}
Consequently, we have proved the following:

\vspace{5mm}
\noindent \textbf{Theorem~3:} All 3-dimensional Lorentzian geometries admitting a geodesic null vector field~$\mathbf{k}$ can be parametrized by \emph{canonical coordinates} ${\{r,\,u,\,x\}}$ in which the metric takes the form
\begin{equation}
\dd s^2 = g_{xx}(r,u,x)\, \dd x^2+2\,g_{ux}(r,u,x)\, \dd u\, \dd x -2\,\dd u\,\dd r+g_{uu}(r,u,x)\, \dd u^2 \,, \label{general nontwist}
\end{equation}
and in which ${\mathbf{k}=\mathbf{\partial}_r\,}$ holds.
\vspace{5mm}

In these canonical coordinates, we can now  explicitly express the null triad frame ${\left\{\mathbf{k},\,\mathbf{l},\,\mathbf{m}\right\}}$ introduced in Eq.~(\ref{triad}). The most natural choice is
\begin{equation}
\mathbf{k}=\partial_r\,, \qquad
\mathbf{l}={\textstyle \frac{1}{2}}\,g_{uu}\,\partial_r+\partial_u \,, \qquad \mathbf{m}=m^x\big(g_{ux}\,\partial_r+\partial_x\big) \,,
\label{triadexpl}
\end{equation}
where ${m^x=1/\sqrt{g_{xx}}}$ to satisfy the normalization condition ${\mathbf{m}\cdot\mathbf{m}=1}$.
It is noted that $\mathbf{l}$ and $\mathbf{m}$ are not geodesics in general.
A direct calculation for the metric (\ref{general nontwist}) immediately shows that the covariant derivative of $\mathbf{k}$ is  ${k_{a;b}=\Gamma^{u}_{ab}=\frac{1}{2}g_{ab,r}}$. In particular, ${k_{r;b}=0=k_{a;r}}$ holds. Explicit form of the \emph{expansion scalar} thus becomes
\begin{equation}
\Theta=\rho=k_{x;x}\,m^x m^x = \frac{g_{xx,r}}{2\,g_{xx}} \,, \label{OptMat}
\end{equation}
implying the relation
\begin{equation}
g_{xx,r}=2\Theta\, g_{xx} \,.
\label{shearfree condition}
\end{equation}

The Christoffel symbols and all coordinate components of the Riemann and Ricci curvature tensors for the general metric (\ref{general nontwist}) are calculated, using Eq.~(\ref{shearfree condition}), in Appendix~A.

Moreover, the expression (\ref{shearfree condition}) can be integrated as
\begin{equation}
{\textstyle g_{xx}=\mathrm{R}^2(r,u,x)\,h_{xx}(u,x) \,, \qquad \hbox{where} \qquad \mathrm{R}=\exp\big(\int\Theta(r,u,x)\,\dd r\big)} \,. \label{IntShearFreeCond}
\end{equation}
The generic case ${\Theta\neq 0}$ gives the expanding \emph{Robinson--Trautman class}. When the expansion vanishes, ${\Theta=0}$, this effectively reduces to ${\mathrm{R}=1}$ so that the spatial metric ${g_{xx}(u,x)}$ is independent of the coordinate $r$. It yields exactly the \emph{Kundt class} of non-expanding, twist-free and shear-free geometries \cite{Stephani:2003,GriffithsPodolsky:2009,PodolskyZofka:2009,OrtaggioPravdaPravdova:2013}. These results can be summarized as:

\vspace{5mm}
\noindent \textbf{Theorem~4:} All geometries of Theorem~3 split into \emph{two distinct classes}. The first one is the \emph{Robinson--Trautman} class of metrics for which the geodesic null vector field~${\mathbf{k}=\mathbf{\partial}_r\,}$  is expanding, ${\Theta\not=0}$, while the second one is the non-expanding \emph{Kundt} class, ${\Theta=0}$. For the Kundt class, the metric coefficient $g_{xx}(u,x)$ in the canonical coordinates (\ref{general nontwist}) is independent of the affine parameter~$r$, while for the Robinson--Trautman subclass, ${g_{xx}(r,u,x)}$ is given by Eq.~(\ref{IntShearFreeCond}) in terms of~$\Theta$.

\subsection{Einstein's field equations: $\Lambda$-vacuum, pure radiation, or gyratons}
\label{Einstein}

Having thus identified all 3-dimensional Lorentzian geometries (admitting a geodesic null vector field), we can now apply the field equations.
Einstein's equations for the metric $g_{ab}$ have the form ${R_{ab}-\frac{1}{2}R\,g_{ab}+\Lambda\, g_{ab}=8\pi\, T_{ab}}$, where a non-vanishing \emph{cosmological constant} $\Lambda$ is admitted, and an arbitrary matter field is given by its energy momentum-tensor $T_{ab}$ with the trace ${T\equiv g^{ab}\,T_{ab}}$. By substituting their trace ${R=2(3\Lambda-8\pi\, T)}$ we obtain
\begin{equation}
R_{ab}= {\textstyle 2\Lambda\,g_{ab}+8\pi\big(T_{ab}-T\,g_{ab}\big)} \,. \label{EinstinEq}
\end{equation}

Our main aim here is to completely integrate the field equations (\ref{EinstinEq}) for the most general metric (\ref{general nontwist}), in both subcases of expanding Robinson--Trautman spacetimes and non-expanding Kundt spacetimes. We will assume a \emph{null radiation matter field aligned} with the geodesic null congruence generated by  ${\mathbf{k}=\mathbf{\partial}_r\,}$. In addition, we will also admit a \emph{gyratonic matter}, which is a generalization of the null radiation field to include a spin of the null source \cite{Bonnor:1970b,FrolovFursaev:2005,KrtousPodolskyZelnikovKadlecova:2012}.

Specifically, we will assume that the only non-vanishing components of the energy-momen\-tum tensor~$T_{ab}$ are ${T_{uu}}$, corresponding to the classical null radiation component, and ${T_{ux}}$, which encodes the inner gyratonic angular momentum. These can be functions of all coordinates:
\begin{equation}
 T_{uu}(r,u,x) \,,\qquad
 T_{ux}(r,u,x)\,. \label{TuuTux}
\end{equation}
We immediately observe from Eq.~(\ref{general nontwist}) that the trace of such an energy-momentum tensor vanishes,
\begin{equation}
 T=0  \quad \Rightarrow \quad R=6\Lambda \,. \label {T=0}
\end{equation}
Moreover, the \emph{conservation of energy-momentum} ${T_{ab;c}\,g^{bc}=0}$ (which follows from the  simplified Bianchi identities ${{R_{ab}}^{;b}=0}$) yields the constraints
\begin{eqnarray}
T_{ax;x}\,g^{xx}+(T_{ar;x}+T_{ax;r})\,g^{rx}+T_{ar;r}\,g^{rr}-T_{ar;u}-T_{au;r}=0\,.
\label{Bianchi1}
\end{eqnarray}
Using the Christoffel symbols (\ref{ChristoffelBegin})--(\ref{ChristoffelEnd}), the non-vanishing terms ${T_{ab;c}}$ entering Eq.~(\ref{Bianchi1}) are
\begin{eqnarray}
T_{xx;x} \rovno -2\Theta\, g_{xx}\,T_{ux}\,,\\
T_{ux;x} \rovno T_{ux,x}+{\textstyle(\Theta\, g^{rx}g_{xx} -\frac{1}{2}g_{ux,r}-\frac{1}{2}g^{xx}g_{xx,x})}\,T_{ux}-\Theta\, g_{xx}\,T_{uu}\,,\\
T_{ur;x} \rovno -\Theta\,T_{ux}\,,\\
T_{ux;r} \rovno T_{ux,r}-\Theta\,T_{ux}\,,\\
T_{ur;u} \rovno -{\textstyle\frac{1}{2}}g^{xx}g_{ux,r}\,T_{ux}\,,\\
T_{uu;r} \rovno T_{uu,r}-g^{xx}g_{ux,r}\,T_{ux}\,.
\end{eqnarray}
For ${a=x}$ and ${a=u}$, respectively, we thus obtain the following two constraints
\begin{eqnarray}
&& T_{ux,r}+\Theta\,T_{ux}=0 \,, \label{EqTup} \\
&& {\textstyle
T_{uu,r}+\Theta\, T_{uu}=g^{xx}\big[\,T_{ux,x}+\big(g_{ux,r}-2\Theta\, g_{ux}-\frac{1}{2}g^{xx}g_{xx,x}\big)T_{ux}\big]
} \label{EqTuu} \,,
\end{eqnarray}
while the constraint for ${a=r}$ is identically satisfied.

Recall that such an energy-momentum tensor represents a generic (aligned) \emph{pure radiation (null) gyratonic matter field}. Of course, by setting ${T_{ux}=0}$ we recover the standard \emph{pure radiation (null dust) without gyratons}, and for ${T_{uu}=0=T_{ux}}$ \emph{vacuum} Einstein's equations are obtained.

Explicit integration of the Einstein field equations (\ref{EinstinEq}) in 2+1 dimensions with the matter field (\ref{TuuTux}) constrained by Eqs.~(\ref{EqTup}) and (\ref{EqTuu}) has to be done separately for the Robinson--Trautman class ${(\Theta\neq 0)}$ and for the Kundt class ${(\Theta=0)}$ of spacetimes. Such a step-by-step integration is performed in Appendices~B and C, respectively.  In subsequent sections~\ref{sec:discussionRT} and \ref{sec:discussionKundt}, we present the summary of the results, together with their physical discussions.

\subsection{Energy conditions}

In the spacetime~(\ref{general nontwist}), a set of three vectors $({\bf k},{\bf l},{\bf m})$ given by Eq.~(\ref{triadexpl}) forms a pseudo-orthonormal basis
\begin{equation}
{\bar E}^a_{(i)}\equiv({\bar E}^a_{(0)},{\bar E}^a_{(1)},{\bar E}^a_{(2)})=({\bf k},{\bf l},{\bf m})\,, \label{psude-bases}
\end{equation}
which satisfies
\begin{equation}
{\bar E}^a_{(i)}{\bar E}_{(j)a}={\bar \eta}_{(i)(j)}=\left(
\begin{array}{ccc}
 0  & -1 & 0 \\
 -1 & 0 & 0 \\
 0 & 0 & 1 \\
\end{array}
\right),
\end{equation}
where ${\bar \eta}_{(i)(j)}$ is the metric in the local Lorentz frame.
The spacetime metric $g_{ab}$ is then given by
\begin{equation}
g_{ab}={\bar \eta}_{(i)(j)}\,{\bar E}^{(i)}_{a}{\bar E}^{(j)}_{b}=-k_al_b-l_ak_b+m_am_b\,.
\end{equation}

From ${\bf k}$ and ${\bf l}$, we construct unit timelike and spacelike vectors ${\bf u}$ and ${\bf s}$ such that
\begin{align}
{\bf u}\equiv &{\textstyle \frac{1}{\sqrt{2}}({\bf k}+{\bf l})=\frac{1}{\sqrt{2}}\bigl[(1+\frac{1}{2}\,g_{uu})\,\partial_r+\partial_u\bigl]}\,,\\
{\bf s}\equiv & {\textstyle\frac{1}{\sqrt{2}}({\bf k}-{\bf l})=\frac{1}{\sqrt{2}}\bigl[(1-\frac{1}{2}\,g_{uu})\,\partial_r-\partial_u\bigl]}\,,
\end{align}
which satisfy ${\bf u}\cdot {\bf u}=-1$, ${\bf s}\cdot{\bf s}=1$, ${\bf u}\cdot{\bf s}=0$, and ${\bf m}\cdot {\bf u}={\bf m}\cdot {\bf s}=0$.
A set of three vectors $({\bf u},{\bf s},{\bf m})$ forms an orthonormal basis in the spacetime (\ref{general nontwist}):
\begin{equation}
{E}^a_{(i)}\equiv ({E}^a_{(0)},{E}^a_{(1)},{E}^a_{(2)})=({\bf u},{\bf s},{\bf m})\,.\label{normal-bases}
\end{equation}
These basis vectors satisfy
\begin{equation}
{E}^a_{(i)}{E}_{(j)a}={\eta}_{(i)(j)}=\left(
\begin{array}{ccc}
 -1  & 0 & 0 \\
 0 & 1 & 0 \\
 0 & 0 & 1 \\
\end{array}
\right)
\end{equation}
and the metric $g_{ab}$ is given by
\begin{equation}
g_{ab}={\eta}_{(i)(j)}{E}^{(i)}_{a}{E}^{(j)}_{b}=-u_au_b+s_as_b+m_am_b \,.
\end{equation}

The components of the energy-momentum tensor $T_{ab}$ in the orthonormal frame (\ref{normal-bases}) are given~by
\begin{equation}
T^{(i)(j)}=\eta^{(i)(k)}\eta^{(j)(l)}E^a_{(k)}E^b_{(l)}\,T_{ab}
=\left(
\begin{array}{ccc}
 \sigma+\nu  & \nu & \zeta \\
 \nu & -\sigma+\nu & \zeta \\
  \zeta & \zeta & -\sigma \\
\end{array}
\right)\,, \label{T(a)(b)}
\end{equation}
where
\begin{equation}
\textstyle{
\sigma=0\,,\qquad \nu=\frac12\, T_{uu}\,,\qquad \zeta=-\frac{1}{\sqrt{2}}\,m^x\,T_{ux}
}\,.
\end{equation}

The expression (\ref{T(a)(b)}) with non-zero $\zeta$ is the 3-dimensional version of the type III energy-momentum tensor in the Hawking--Ellis classification~\cite{Hawking:1973uf,mmv2017}, and reduces to a special case of the type II if $\zeta\equiv 0$ holds.
The type III energy-momentum tensor violates the null energy condition and therefore the weak and dominant energy conditions as well~\cite{energyconditions}.

In the case of ${\zeta\equiv 0}$ (that is for ${T_{ux}=0}$), as shown in~\cite{energyconditions}, the standard energy conditions are equivalent to
\begin{itemize}
\item \emph{Null} energy condition (NEC): ${\nu\ge 0}$\,,
\item \emph{Weak} energy condition (WEC): ${\nu\ge 0}$ and ${\sigma\ge 0}$\,,
\item \emph{Strong} energy condition (SEC): ${\nu\ge 0}$ and ${\sigma\le 0}$\,,
\item \emph{Dominant} energy condition (DEC): ${\nu\ge 0}$ and ${\sigma\ge 0}$\,.
\end{itemize}
Thus, while gyratons ($T_{ux}\ne 0$) violate NEC (and hence WEC and DEC as well), a pure radiation ($T_{ux}\equiv 0$) respects all the standard energy conditions if and only if $T_{uu}\ge 0$.

\section{All Robinson--Trautman solutions and their properties}
\label{sec:discussionRT}

The complete integration of Einstein's field equations for a general  3-dimensional Robinson--Trautman metric in vacuum, with a cosmological constant $\Lambda$, and possibly a pure radiation field ${T_{uu}}$ and gyratons ${T_{ux}}$ is performed in appendix~B. Equations~(\ref{ExplTuu}) and (\ref{ExplTup}) show that the matter field necessarily takes the form
\begin{eqnarray}
&& T_{uu} = \frac{\mathcal{N}}{r}-\frac{P(P\mathcal{J})_{,x}}{r^2}+\frac{fP^2\mathcal{J}}{r^3} \,, \label{RTTuu}\\
&& T_{ux} = \frac{\mathcal{J}}{r} \,, \label{RTTup}
\end{eqnarray}
where ${\mathcal{N}(u,x)}$ and ${\mathcal{J}(u,x)}$ are functions determining the (density of) energy and angular momentum. Notice that ${T_{uu}\to0}$ and ${T_{ux}\to0}$ as ${r\to \infty}$, i.e., asymptotically the solutions ``become vacuum''. Indeed, it can be shown that ${R_{abcd}\to \Lambda\,(g_{ac}\,g_{bd}-g_{ad}\,g_{bc})}$ in the limit of ${r\to \infty}$ with constant $u$ and $x$, namely the spacetime is asymptotically maximally symmetric (Minkowski, dS, or AdS), at least locally.

The corresponding generic metric can be written in the form (\ref{RTmetric}), that is
\begin{eqnarray}
\dd s^2 \rovno \frac{r^2}{P^2}\, \dd x^2+2\,(e\,r^2+f\,)\,\dd u \dd x -2\,\dd u\dd r \nonumber \\
&& +\Big(-a  +2\big[ P(Pe)_{,x}+(\ln P)_{,u} \big]\,r +(\Lambda+P^2e^2)\,r^2\Big)\, \dd u^2 \,. \label{RTmetricsummary}
\end{eqnarray}
The metric functions $P(u,x)$, $e(u,x)$, $f(u,x)$ and $a(u,x)$ are constrained just by two field equations (\ref{fieleq_ux}) and (\ref{RTEq2}):
\begin{eqnarray}
 a_{,x} \rovno  c f - 2 f_{,u}-16\pi\, \mathcal{J} \,, \label{RTE1}\\
 a_{,u}\rovno ac+\triangle c+2(\Lambda+P^2e^2)P(Pf)_{,x} +3P^2f(P^2e^2)_{,x}  \nonumber\\
&& -2P^2f\,e_{,u} -P^2e(4 f_{,u}-cf+48\pi\, \mathcal{J})
+ 16\pi\,\mathcal{N} \,, \label{RTE2}
\end{eqnarray}
where ${\triangle c \equiv P(Pc_{,x})_{,x}}$
is the transverse-space Laplace operator applied on the function $c$, defined by $c \equiv 2\big[P(Pe)_{,x}+(\ln P)_{,u}\big]$. (See Eqs.~(\ref{Laplace}) and (\ref{idno3}).)

Generically, by prescribing an \emph{arbitrary gyratonic function} $\mathcal{J}$ (as well as \emph{any} metric functions ${P, e, f}$) we can always integrate (\ref{RTE1}) to obtain ${a(u,x)}$. Subsequently, its partial derivative $a_{,u}$ (and other given functions) uniquely determines the pure radiation energy profile $\mathcal{N}$ via the field equation (\ref{RTE2}).

It is remarkable that in ${D=3}$ the function $f(u,x)$ in the metric (\ref{RTmetricsummary}) \emph{remains arbitrary} and, in general, \emph{non-vanishing}. This is an \emph{entirely new feature which does not occur in dimensions} ${D\ge4}$. Indeed, it was demonstrated in \cite{PodOrt06,OrtPodZof08,OrtaggioPodolskyZofka:2015} that for the Robinson--Trautman class of spacetimes in four and any higher dimensions necessarily ${f_i=0}$ for all ${(D-2)}$ spatial components. In this sense, the ${D=3}$ case is richer than the ${D\ge4}$ cases.

\subsection{Gauge freedom}
\label{gaugeRT}

The most general form of the Robinson--Trautman metric (\ref{RTmetricsummary}) can be simplified using the gauge freedom. In the following, we consider the coordinate transformations ${(u,x)\to (u,x')}$ such that ${x=x(u,x')}$.
Then, we have
\begin{eqnarray}
&&\dd x=\frac{\partial x}{\partial u}\dd u+\frac{\partial x}{\partial x'}\dd x'\,.
\end{eqnarray}
Since this equation gives
\begin{eqnarray}
&&\dd x=\frac{\partial x}{\partial u}\dd u+\frac{\partial x}{\partial x'}\biggl(\frac{\partial x'}{\partial u}\dd u+\frac{\partial x'}{\partial x}\dd x\biggl)=\biggl(\frac{\partial x}{\partial u}+\frac{\partial x}{\partial x'}\frac{\partial x'}{\partial u}\biggl)\dd u+\frac{\partial x}{\partial x'}\frac{\partial x'}{\partial x}\dd x\, ,
\end{eqnarray}
we have the following relations:
\begin{eqnarray}
\frac{\partial x'}{\partial u}=-\frac{\partial x}{\partial u}\biggl(\frac{\partial x}{\partial x'}\biggl)^{-1},\qquad \frac{\partial x'}{\partial x}=\biggl(\frac{\partial x}{\partial x'}\biggl)^{-1}\, ,
\end{eqnarray}
with which we obtain
\begin{eqnarray}
&&\frac{\partial}{\partial u}\biggl|_{(u,x)}=\biggl(\frac{\partial}{\partial u}+\frac{\partial x'}{\partial u}\frac{\partial}{\partial x'}\biggl)\biggl|_{(u,x')}=\biggl[\frac{\partial}{\partial u}-\frac{\partial x}{\partial u}\biggl(\frac{\partial x}{\partial x'}\biggl)^{-1}\frac{\partial}{\partial x'}\biggl]\biggl|_{(u,x')}\,,\\
&&\frac{\partial}{\partial x}\biggl|_{(u,x)}=\frac{\partial x'}{\partial x}\frac{\partial}{\partial x'}\biggl|_{(u,x')}=\biggl(\frac{\partial x}{\partial x'}\biggl)^{-1}\frac{\partial}{\partial x'}\biggl|_{(u,x')}\,.
\end{eqnarray}
Consequently, the Robinson--Trautman metric (\ref{RTmetricsummary}) is transformed to be
\begin{eqnarray}
\dd s^2 \rovno \frac{r^2}{P^2}\biggl(\frac{\partial x}{\partial x'}\biggl)^2\dd {x'}^2+2\frac{\partial x}{\partial x'}\biggl(\frac{r^2}{P^2}\frac{\partial x}{\partial u}+e\,r^2+f\biggl)\dd u\dd x' -2\,\dd u\dd r \nonumber \\
&& +\Bigg[-a  +2\biggl[ P\biggl(\frac{\partial x}{\partial x'}\biggl)^{-1}(Pe)_{,x'}+(\ln P)_{,u}-\frac{\partial x}{\partial u}\biggl(\frac{\partial x}{\partial x'}\biggl)^{-1}(\ln P)_{,x'} \biggl]\,r \nonumber \\
&&\qquad +(\Lambda+P^2e^2)\,r^2+\frac{r^2}{P^2}\biggl(\frac{\partial x}{\partial u}\biggl)^2+2\,(e\,r^2+f\,)\frac{\partial x}{\partial u}\Bigg]\, \dd u^2 \,,
\end{eqnarray}
where we have assumed $\frac{\partial x}{\partial x'}\ne 0$.
Indeed, there are two natural gauge choices for the Robinson--Trautman metric (\ref{RTmetricsummary}):

\begin{itemize}
\item \textbf{Setting ${e=0}$}

The function $e(u,x)$, introduced in the off-diagonal metric component ${g_{ux}=e\,r^2+f}$, \emph{can locally be removed}\footnote{But it may be \emph{convenient to retain} the function $e$ non-trivial because it may directly encode some physical or geometrical property, e.g. the gyratonic aspect of the matter (or as in the Garc\'{\i}a D\'{\i}az--Pleba\'nski 1981 gauge form \cite{GarciaPlebanski:1981,BicakPodolsky:1999a} of the ${D=4}$ Robinson--Trautman metric, where it directly encodes amplitudes of gravitational waves).} by a gauge transformation ${x \to x'}$ such that
\begin{equation}
x = -{\textstyle\int} P^2(u,x')\,e(u,x')\,\dd u \,. \label{gauge of e}
\end{equation}
(In the degenerate case when  ${\frac{\partial x}{\partial x'}=0}$, namely, if both $P$ and $e$ are independent of $x$, we remove $e$ by ${x= -\int P^2(u)e(u)\,\dd u - \lambda\,x'}$, where $\lambda$ is a suitable real parameter \cite{PodOrt06}.) After this coordinate transformation and redefinitions such that ${P/\frac{\partial x}{\partial x'}\to P}$, ${f\frac{\partial x}{\partial x'}\to f}$, and ${a+2P^2ef \to a}$, the metric (\ref{RTmetricsummary}) reduces to (without a prime on the new coordinate $x'$)
\begin{equation}
\dd s^2 = \frac{r^2}{P^2}\, \dd x^2+2\,f\,\dd u \dd x -2\,\dd u\dd r
+\Big[-a + 2\,(\ln P)_{,u}\,r + \Lambda\,r^2\,\Big]\, \dd u^2 \,, \label{RTmet}
\end{equation}
and the remaining two field equations (\ref{RTE1}), (\ref{RTE2}) are simplified considerably as
\begin{eqnarray}
a_{,x} \rovno 2\,\big[ f\,(\ln P)_{,u} - f_{,u} \big] -16\pi\, \mathcal{J} \,, \label{RTEq1eje0}\\
a_{,u}\rovno 2\,a(\ln P)_{,u}+2\,\triangle (\ln P)_{,u}+2\Lambda\,P(Pf)_{,x} + 16\pi\,\mathcal{N} \,, \label{RTEq2eje0}
\end{eqnarray}
where
\begin{equation}
\triangle (\ln P)_{,u} \equiv P[P(\ln P)_{,ux}]_{,x}\,.
\label{Laplace2}
\end{equation}
Prescribing \emph{arbitrary} functions $P(u,x)$, $f(u,x)$ and $\mathcal{J}(u,x)$, we can thus  integrate equation (\ref{RTEq1eje0})  to obtain $a(u,x)$. The second equation (\ref{RTEq2eje0}) then determines the corresponding unique form of the pure radiation profile $\mathcal{N}(u,x)$.

\item \textbf{Setting ${P=1}$}

Alternatively, it is possible to employ the gauge freedom to simplify the metric function $P$ to ${P=1}$, at the expense of keeping the off-diagonal metric function $e(u,x)$. This is achieved by the transformation ${x \to x'}$ given by
\begin{equation}
x = {\textstyle\int} P(u,x')\,\dd x' \,, \label{gauge of P}
\end{equation}
and redefinitions ${P^{-1}\frac{\partial x}{\partial u}+Pe\to e}$, ${Pf\to f}$, and ${a-2\frac{\partial x}{\partial u}f\to a}$. This leads to the metric
\begin{equation}
\dd s^2 = r^2\, \dd x^2+2\,(e\,r^2+f\,)\,\dd u \dd x -2\,\dd u\dd r +\Big[-a  +2\,e_{,x}\,r +(\Lambda+e^2)\,r^2\Big]\, \dd u^2 \,, \label{RTmetP=1}
\end{equation}
for which the field equations (\ref{RTE1}), (\ref{RTE2}) are
\begin{eqnarray}
 a_{,x} \rovno  c f - 2 f_{,u}-16\pi\, \mathcal{J} \,, \label{RTE1Pje1}\\
 a_{,u}\rovno ac+c_{,xx}+2(\Lambda+e^2)f_{,x} +3f(e^2)_{,x}
  -2f\,e_{,u} -e(4 f_{,u}-cf+48\pi\, \mathcal{J})
+ 16\pi\,\mathcal{N} \,, \label{RTE2Pje1}
\end{eqnarray}
where ${c \equiv 2e_{,x}}$. These equations look more complicated  than Eqs.~(\ref{RTEq1eje0}) and (\ref{RTEq2eje0}).

\end{itemize}

In any case, this is a very large family of explicit exact solutions to ${D=3}$ Einstein's gravity. In particular, as a unique feature, it admits the non-trivial off-diagonal metric function $f(u,x)$ which gives an additional degree of freedom.

\subsection{Special subcase ${f=0}$}

In this special subcase, the Robinson--Trautman metric (in the gauge ${e=0}$ corresponding to Eq.~(\ref{RTmet})) is further simplified to
\begin{equation}
\dd s^2 = \frac{r^2}{P^2}\, \dd x^2 -2\,\dd u\dd r
+\Big[-a + 2\,(\ln P)_{,u}\,r + \Lambda\,r^2\,\Big]\, \dd u^2 \,, \label{RTmet_f=0}
\end{equation}
with $a(u,x)$ such that
\begin{eqnarray}
a_{,x} \rovno -16\pi\, \mathcal{J} \,, \label{RTEq1eje0_f=0}\\
a_{,u} - 2\,a(\ln P)_{,u}-2\,\triangle (\ln P)_{,u} \rovno \ \  16\pi\,\mathcal{N} \,. \label{RTEq2eje0_f=0}
\end{eqnarray}
Notice that these field equations are \emph{fully independent of the cosmological constant} $\Lambda$. We are free to prescribe arbitrary metric functions $P(u,x)$ and ${a(u,x)}$. The corresponding matter contents $\mathcal{J} $ and $\mathcal{N}$ are then evaluated by Eqs.~(\ref{RTEq1eje0_f=0}) and (\ref{RTEq2eje0_f=0}), respectively.

Interestingly, due to Eq.~(\ref{RTEq1eje0_f=0}), the key metric function $a(u)$ \emph{does not depend on the spatial coordinate $x$ if, and only if}, ${\mathcal{J}=0}$, i.e., \emph{in the absence of gyratons}. The remaining field equation (\ref{RTEq2eje0_f=0}) couples the metric functions $a$ and $(\ln P)_{,u}$ to the matter profile $\mathcal{N}(u,x)$.

There also exists a special gyratonic solution given by ${P=1}$. In such a case, the metric has a very simple form
\begin{equation}
\dd s^2 = r^2\, \dd x^2 -2\,\dd u\dd r
+\big[-a + \Lambda\,r^2\,\big]\, \dd u^2 \,, \label{RTmet_f=0_P=1}
\end{equation}
where
\begin{eqnarray}
a_{,x} = -16\pi\,\mathcal{J} \,, \qquad
a_{,u} =  16\pi\,\mathcal{N} \,. \label{RTEq2eje0_f=0_P=1}
\end{eqnarray}
These two field equations are consistent if, and only if, ${\mathcal{N}_{,x}=-\mathcal{J}_{,u}}$. In fact, this solution can be interpreted (replacing $x$ by $\phi$ with a compact range) as a \emph{Vaidya-type object} emitting a null radiation gyratonic matter, i.e., a gyratonic generalization of the solution given by Eqs.~(\ref{Vaidya}) and (\ref{RTEVaidya}).

\subsection{Arbitrarily moving object with pure radiation (${\mathcal{J}=0}$)}
\label{sec_moving object}

It is interesting to observe that the Robinson--Trautman line element (\ref{RTmet_f=0}) with only pure null radiation matter field  ${T_{uu}=\mathcal{N}\,r^{-1}}$, where $\mathcal{N}(u,x)$ is given by expression (\ref{RTEq2eje0_f=0}), is an \emph{exact solution of Einstein's equations representing an arbitrarily moving object}. It accelerates due to an \emph{anisotropic emission} of the null matter. In fact, it is  the 3-dimensional version of the so-called \emph{Kinnerley photon rocket} moving in Minkowski, de~Sitter, or anti--de~Sitter universe~\cite{Kinnersley:1969,Bonnor:1994,DainMoreschiGleiser:1996,Podolsky:2008,GursesSarioglu:2002,Podolsky:2011,NewmanUnti:1963}.

To describe and physically interpret this family of spacetimes, it is better to introduce \emph{angular coordinate} ${\phi\in[0,2\pi)}$ instead of ${x\in {\rm R}}$ (i.e., to compactify the transverse space to a circle~$S^1$). Moreover, for convenience we relabel $a(u)$ to ${-\mu(u)}$ in the metric (\ref{RTmet_f=0}),
\begin{equation}
\dd s^2 = \frac{r^2}{P^2}\, \dd \phi^2 -2\,\dd u\dd r
+\Big[\,\mu(u) + 2\,(\ln P)_{,u}\,r + \Lambda\,r^2\,\Big]\, \dd u^2 \,, \label{photon_rocket}
\end{equation}
and we take the key function $P(u, \phi)$ to be of the following special form
\begin{equation}
P(u, \phi) = \dot{z}^0(u)-\dot{z}^1(u)\cos\phi-\dot{z}^2(u)\sin\phi \,, \label{p_for_photon_rocket}
\end{equation}
where ${\dot{z}^0(u), \dot{z}^1(u), \dot{z}^2(u)}$ are \emph{arbitrary functions of the coordinate}~$u$ that satisfy a normalization condition ${(\dot{z}^0)^2-(\dot{z}^1)^2-(\dot{z}^2)^2=1}$ and a dot denotes derivative with respect to $u$. Such a constraint is always satisfied if we employ a parametrization ${\dot{z}^0\equiv \sqrt{1+v^2}}$, ${\dot{z}^1\equiv v\sin\psi}$, ${\dot{z}^2\equiv v\cos\psi}$,  that is
\begin{equation}
P(u, \phi) = \sqrt{1+v^2(u)}-v(u)\sin\psi(u)\cos\phi-v(u)\cos\psi(u)\sin\phi \, . \label{p_for_photon_rocket_conv}
\end{equation}

\begin{figure}[htbp]
\begin{center}
\includegraphics[width=0.6\linewidth]{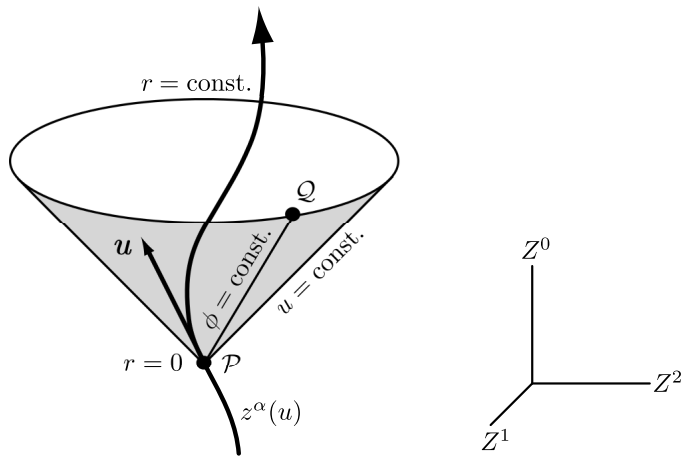}
\caption{\label{fig1}
The coordinate system (\ref{photon_rocket}) is adapted to an arbitrarily moving test particle in a space with Minkowski coordinates ${Z^0, Z^1, Z^2}$. At any event ${\cal P}$ of its worldline ${Z^\alpha=z^\alpha(u)}$, with velocity {\boldmath$u$}, a future null cone is labeled by $u$ and it is assigned the value of the corresponding proper time. The coordinate $r$ is an affine parameter along null geodesics which connect ${\cal P}$ with any event ${\cal Q}$ on the cone. The position of~${\cal Q}$ on the circle $u=\hbox{const.}$, $r=\hbox{const.}$ is specified by the angular coordinate $\phi$. The moving particle is always located at ${r=0}$.
}
\end{center}
\end{figure}

In fact, ${\{\dot{z}^0(u), \dot{z}^1(u), \dot{z}^2(u)\}}$ are \emph{components of the 3-velocity} {\boldmath$u$} of an object whose \emph{mass} is proportional to the function $\mu(u)$, located at the (singular) origin ${r=0}$ of the comoving coordinates~\cite{NewmanUnti:1963,Podolsky:2011}. This is illustrated in Fig.~\ref{fig1}. The function $v(u)$ in Eq.~(\ref{p_for_photon_rocket_conv}) thus denotes the \emph{speed} of the object in two spatial directions, while the angle $\psi(u)$ determines the actual \emph{direction} of the spatial velocity at an instant~$u$. Since these three functions can be prescribed arbitrarily, the metric (\ref{photon_rocket}) with the function (\ref{p_for_photon_rocket}) or (\ref{p_for_photon_rocket_conv}) is an \emph{exact solution for the gravitational field of an arbitrarily moving massive object}.

Notice that, compared to the previously studied cases in ${D\ge 4}$  \cite{Podolsky:2008,Podolsky:2011,NewmanUnti:1963}, there is no Gaussian/Ricci curvature term ${K}$ in the component $g_{uu}$ of the metric (\ref{photon_rocket}) because the 1-dimensional transverse space spanned by $\phi$ is always flat and thus ${K=0}$.

The single field equation (\ref{RTEq2eje0_f=0}) determines the corresponding $\mathcal{N}$ as
\begin{eqnarray}
\mathcal{N}(u,\phi) \rovno
   {\textstyle\frac{1}{16\pi}}\Big[-\mu_{,u}+2\mu\,(\ln P)_{,u}-2\,\triangle (\ln P)_{,u} \Big]\nonumber\\
\rovno {\textstyle\frac{1}{8\pi}}\Big[-{\textstyle\frac{1}{2}}\mu_{,u}+\big(\,\mu-(P_{,\phi})^2  \big)\,(\ln P)_{,u}-P P_{,u\phi\phi}+P_{,u} P_{,\phi\phi}+P_{,u\phi} P_{,\phi}  \Big]\,, \label{NKinnersley}
\end{eqnarray}
where we have also employed the relation (\ref{Laplace2}).
Interestingly, for the specific function (\ref{p_for_photon_rocket_conv}) the term ${\triangle (\ln P)_{,u}}$ is just equal to ${-(\ln P)_{,u}}$, so that the null matter field is simply given by
\begin{equation}
\mathcal{N}(u,\phi) =
   {\textstyle\frac{1}{8\pi}}\Big[-{\textstyle\frac{1}{2}}\mu_{,u}+(\mu+1)(\ln P)_{,u} \Big]\,. \label{NKinnersleySimple}
\end{equation}
Notice that the motion of the object with ${\mathcal{N}>0}$ is related to its \emph{decreasing mass function} $\mu(u)$.

As an interesting special subcase, we get an explicit solution representing \emph{radial straight flight} of the photon rocket. It is obtained, e.g.,  by setting ${\psi(u)=\frac{\pi}{2}}$, so that ${\dot{z}^1(u)=v(u)}$ and ${\dot{z}^2(u)=0}$, and hence ${P = \sqrt{1+v^2(u)}-v(u)
\cos\phi }$. In fact, it is convenient to express the speed function as
\begin{equation}
v(u) = \dot{z}^1(u) = \sinh\big({\textstyle\int}\alpha(u)\,\dd u\big)\,,
 \label{acceleration function}
\end{equation}
where the function $\alpha(u)$ determines the instantaneous \emph{acceleration} of the rocket. The key function then reads
\begin{equation}
P = \cosh\big({\textstyle\int}\alpha\,\dd u\big)-\sinh\big({\textstyle\int}\alpha\,\dd u\big)\, \cos\phi\,.
 \label{transfX-P}
\end{equation}
Now we perform the transformation
\begin{equation}
\sin\varphi =\frac{\epsilon\sin\phi}{P(u,\phi)}\,,
 \label{transfX}
\end{equation}
where $\epsilon=\pm1$.
Equations~(\ref{transfX-P}) and (\ref{transfX}) imply
\begin{equation}
\cos\varphi=P^{-1}[\sinh \big({\textstyle\int}\alpha\,\dd u\big)-\cosh \big({\textstyle\int}\alpha\,\dd u\big) \cos\phi]\,,
 \label{transfX-cos}
\end{equation}
by using a remaining degree of freedom of transformation $\varphi\to \varphi+\pi$ because the sign of $\epsilon$ in Eq.~(\ref{transfX}) is arbitrary. Equations~(\ref{transfX-P}), (\ref{transfX}), and (\ref{transfX-cos}) also imply interesting relations ${(\ln P)_{,u} = \alpha(u)\,\cos\varphi}$,
${\epsilon\cot(\phi/2)=\tan(\varphi/2)\,\exp\big({\textstyle\int}\alpha(u)\,\dd u\big)}$, and
${\dd \varphi+\alpha(u)\sin\varphi\,\dd u=-\epsilon\:\dd \phi/P}$, so that finally the metric (\ref{photon_rocket}) is put into the following form:
\begin{equation}
\dd s^2 = r^2\big[\,\dd \varphi+\alpha(u)\sin\varphi\,\dd u\,\big]^2 -2\,\dd u\dd r
+\big[\,\mu(u) + 2\,r\,\alpha(u)\cos\varphi  + \Lambda\,r^2\,\big]\, \dd u^2 \,. \label{photon_rocket 1 direction}
\end{equation}
This is the ${D=3}$ version of the classic Kinnersley metric \cite{Kinnersley:1969, Bonnor:1994} with a cosmological constant \cite{Podolsky:2008, Podolsky:2011} and Eq.~(19.54) in \cite{GriffithsPodolsky:2009}. Gravitational field generated by a photon rocket \emph{uniformly accelerating} in a single spatial direction is obtained by simply assuming ${\alpha=\hbox{const.}}$ in the metric (\ref{photon_rocket 1 direction}).

\subsection{Standing Vaidya-type object with pure radiation (${\mathcal{J}=0}$)}
\label{sec_standing object}

As an important special subcase of the family of solutions given by Eqs.~(\ref{photon_rocket}) and (\ref{p_for_photon_rocket}), we immediately obtain an exact gravitational field of a \emph{non-moving object}. Of course, this has a trivial spatial velocity ${v=0}$, equivalent to ${\dot{z}^1(u)=0=\dot{z}^2(u)}$, and the corresponding metric function (\ref{p_for_photon_rocket_conv}) is simply ${P=1}$. The metric (\ref{photon_rocket}) then reduces~to
\begin{equation}
\dd s^2 = r^2\, \dd \phi^2 -2\,\dd u\dd r
+\big[\,\mu(u) + \Lambda\,r^2\,\big]\, \dd u^2 \,, \label{Vaidya}
\end{equation}
and it is circularly symmetric. The corresponding null matter field is given by
\begin{equation}
\mathcal{N}(u) = - {\textstyle\frac{1}{16\pi}}\,\mu_{,u}\,, \label{RTEVaidya}
\end{equation}
and it is also independent of $\phi$. In other words, the radiation from the source is \emph{isotropic} and thus causes no acceleration. The same result is obtain by setting ${\alpha=0}$ in the metric (\ref{photon_rocket 1 direction}).

\subsection{Vacuum solutions: Minkowski, de~Sitter, anti--de~Sitter (${\mathcal{N}=0=\mathcal{J}}$)}
\label{sec_MAdS}

It is well-known that in ${D=3}$, all \emph{vacuum} solutions are just maximally symmetric background spacetimes of constant curvature, that is Minkowski, de~Sitter, anti--de~Sitter universe, according to the sign of the cosmological constant $\Lambda$. In the context of the family of metrics (\ref{RTmet})--(\ref{RTEq2eje0}),
this occurs if, and only if, ${\mathcal{N}=0=\mathcal{J}}$.
In such a case we  need to satisfy a \emph{single integrability condition} for the function $a(x,u)$, namely
\begin{equation}
a_{,xu}=a_{,ux}\,. \label{integrabilit}
\end{equation}
The left-hand side is obtained by taking the $u$-derivative of Eq.~(\ref{RTEq1eje0}), while the right-hand side is obtained from the $x$-derivative of Eq.~(\ref{RTEq2eje0}), yielding the following constraint:
\begin{eqnarray}
a\,(\ln P)_{,ux} \!\!\!&=&\!\!\!  f\,(\ln P)_{,uu} - 2f\,[(\ln P)_{,u}]^2
 +3 f_{,u}\,(\ln P)_{,u}-f_{,uu}   \nonumber\\
&& -[\triangle (\ln P)_{,u}]_{,x}-\Lambda\, [P(Pf)_{,x}]_{,x} \,. \label{integrability}
\end{eqnarray}
There are two distinct subcases to be considered:
\begin{itemize}
\item \textbf{The case} ${(\ln P)_{,ux} \not=0}$ allows us to divide both parts of equation (\ref{integrability}) by the term ${(\ln P)_{,ux}}$. We thus obtain an explicit expression for the function $a(u,x)$, uniquely determined by the remaining free metric functions $P(u,x)$ and $f(u,x)$.

\item \textbf{The case} ${(\ln P)_{,ux} =0 \Leftrightarrow}$ the function $P$ is \emph{separable} in the form ${P(u,x)=U(u)X(x)}$. In such a case, the function $a(u,x)$ is given by equation (\ref{RTEq1eje0}), while the functions $P(u,x)$ and $f(u,x)$ must satisfy the constraint (\ref{integrability}) with vanishing left-hand side. Interestingly, such a condition can be rewritten as
    \begin{equation}
\big[\,U^{-1} (U^{-1}f)_{,u}\big]_{,u}= -\Lambda\, \big[X(Xf)_{,x}\big]_{,x} \,. \label{constraintPf}
\end{equation}
Assuming the function $f$ is \emph{also separable}, this equation can be explicitly solved by separation of variables $u$ and $x$.
Moreover, when ${\Lambda=0}$, a general solution of Eq.~(\ref{constraintPf}) is ${f(u,x)=U(u)\,\big[\alpha(x)\int U(u)\,\dd u+\beta(x)\big]}$, where $\alpha(x)$ and $\beta(x)$ are arbitrary functions.

\end{itemize}

The vacuum Robinson--Trautman metric is \emph{simplified for} ${f=0}$, i.e., for the ${\mathcal{J}=0=\mathcal{N}}$ subcase of the metric (\ref{RTmet_f=0}). In this case, the two metric functions $a(u)$ and ${P(x,u)}$ are related by the single field equation (\ref{RTEq2eje0_f=0}), namely
\begin{equation}
\big[\triangle + a \big](\ln P)_{,u} = {\textstyle \frac{1}{2}}a_{,u}\,. \label{f=0vacuum}
\end{equation}
Prescribing an arbitrary $a(u)$, we obtain $P(x,u)$ by solving this differential equation. \emph{Assuming separability}, ${P(u,x)=U(u)X(x)}$, the only possibility is ${U=\sqrt{a(u)}}$, and (by redefining $x$) the corresponding metric takes the form
\begin{equation}
\dd s^2 = (r^2/a)\, \dd x^2 -2\,\dd u\dd r
+\big[-a + (\ln a)_{,u}\,r + \Lambda\,r^2\,\big]\, \dd u^2 \,. \label{RTmet_f=0vacuum}
\end{equation}
The transformation
\begin{equation}
R=r/\sqrt{a(u)}\,,\qquad U={\textstyle\int}\sqrt{a(u)}\,\dd u\,, \qquad \phi=x\,, \label{transfRTmet_f=0vacuum}
\end{equation}
puts it into the ``Eddington form'' of  maximally symmetric spaces
\begin{equation}
\dd s^2 = R^2\, \dd \phi^2 -2\,\dd U\dd R
+\big(-1 + \Lambda\,R^2\,\big)\, \dd U^2 \,, \label{MinkAdS}
\end{equation}
which is actually the metric (\ref{Vaidya}) with ${\mu=-1}$. The following canonical form of Minkowski and (anti-)de~Sitter space
\begin{equation}
\dd s^2 = R^2\, \dd \phi^2 + (1-\Lambda\,R^2)^{-1}\, \dd R^2
- (1-\Lambda\,R^2)\, \dd T^2 \,, \label{MinkAdScanon}
\end{equation}
is obtained by expressing the null coordinate as ${U=T-\int(1-\Lambda\,R^2)^{-1}\dd R}$.

\section{All Kundt solutions and their properties}
\label{sec:discussionKundt}

In appendix~C, we completely integrate Einstein's field equations for a general  3-dimensional Kundt line element in vacuum, with a cosmological constant $\Lambda$, and possibly a null radiation matter field $T_{uu}$ with a gyratonic component $T_{ux}$. These matter components have a generic form (\ref{KExplTup}) and~(\ref{KExplTuu}):
\begin{eqnarray}
&& T_{uu} = \mathcal{N}+P^2\Big[\mathcal{J}_{,x} + \mathcal{J}\Big(f+ \frac{P_{,x}}{P}\Big) \Big]\,r \,, \label{KundtTuu}\\
&& T_{ux} = \mathcal{J} \,. \label{KundtTup}
\end{eqnarray}
The metric can be written in the form (\ref{Kundtmetric}) which, in general, contains five metric functions $P(u,x)$, $e(u,x)$, $f(u,x)$, $a(u,x)$, and $b(u,x)$. This large family of spacetimes splits into \emph{two distinct subclasses}, namely:

\begin{itemize}
\item \textbf{The case} ${f=0}$ necessarily requires ${\Lambda=0}$ and ${F=0}$, so that such a general metric reads
\begin{eqnarray}
\dd s^2 \rovno \frac{1}{P^2}\, \dd x^2+2\,e\,\dd u \dd x -2\,\dd u\dd r +\big(a+b\,r\big)\, \dd u^2 \,, \label{Kundtmetric_f=0}
\end{eqnarray}
where $a$, $b$, $e$, and $P$ remain \emph{arbitrary functions} of $u$ and $x$. This represents a class of exact solutions of Einstein's equations with a gyrating null matter
characterized by the functions
\begin{eqnarray}
\mathcal{J}(u,x) \rovno -{\textstyle\frac{1}{16\pi}}\,b_{,x}\,, \label{Jf=0sumar}\\
\mathcal{N}(u,x) \rovno {\textstyle\frac{1}{8\pi}}\,P^2 \Big[-{\textstyle\frac{1}{2}}\Big(a_{,xx}+a_{,x}\frac{P_{,x}}{P}\Big)
-{\textstyle\frac{1}{2}}b\Big(e_{,x}+\frac{P_{,x}}{P}\,e+ \frac{P_{,u}}{P^3}\Big) \nonumber\\
&&\hspace{14.0mm}
+e_{,ux}+ \frac{P_{,x}}{P}\,e_{,u}
+ \frac{P_{,uu}}{P^3}-2\frac{P_{,u}^2}{P^4}\Big]+2P^2 e\,\mathcal{J} \,. \label{Nf=0sumar}
\end{eqnarray}
(See Eqs.~(\ref{JKundtf=0}) and (\ref{RuuKundt2f=0}).)
They explicitly relate the matter functions to the metric functions of the gravitational field. While the gyratonic part  $\mathcal{J}$ is directly determined by $b_{,x}$, the null matter energy profile $\mathcal{N}$ is given primarily by the spatial derivatives of $a$.

Whenever $b(u)$ is independent of $x$, the gyrating component vanishes (${\mathcal{J}=0}$) and the spacetimes contains only pure radiation. When ${\mathcal{J}=0=\mathcal{N}}$, the metric (\ref{Kundtmetric_f=0}) is a \emph{vacuum solution}. In fact, it represents Minkowski spacetime. For ${P=1}$, ${e=b=0}$, and ${a=0}$, this is directly the usual form of flat spacetime ${\dd s^2 = \dd x^2-2\,\dd u\dd r}$ in the double-null coordinates (as well as for any $a$ linear in $x$).

\item \textbf{The case} ${f\not=0}$ is the most general form of the Kundt metrics with the matter content studied here,
\begin{eqnarray}
\dd s^2 \rovno \frac{1}{P^2}\, \dd x^2+2\,(e+f\,r)\,\dd u \dd x -2\,\dd u\dd r  +\Big[a+b\,r +\big(\Lambda+{\textstyle\frac{1}{4}} \F \big)\,r^2 \Big]\, \dd u^2 \,, \label{Kundtmetric_fnot=0}
\end{eqnarray}
where ${\F \equiv P^2f^2}$ satisfies Eq.~(\ref{F}), namely
\begin{equation}
 \F_{,x}+ (\F+4\Lambda)f=0 \,. \label{Fsummary}
\end{equation}
Its general solution can be obtained by treating $F(u,x)$ as an arbitrary generating function, from which we obtain ${f=-\F_{,x}/(\F+4\Lambda)}$ and  ${P^2=\F/f^2=\F(\F+4\Lambda)^2/(\F_{,x})^2}$.
Alternatively, we  prescribe any $f(x,u)$, which generates ${\F=-4\Lambda+g(u)\,\exp[-\int\! f(x,u)\,\dd x]}$ and ${P^2=\F/f^2}$.

In this case ${f\not=0}$, the gyratonic matter field is characterized by Eqs.~(\ref{Rux2}) and (\ref{RuuKundt2fnot=0}), namely
\begin{eqnarray}
\mathcal{J}(u,x) \rovno {\textstyle\frac{1}{16\pi} \Big[\, f_{,u}-b_{,x}
 -\frac{1}{2}(F+4\Lambda)e-f\,(\ln P)_{,u}\,\Big]}\,, \label{Jfnot=0sumar}\\
\mathcal{N}(u,x) \rovno {\textstyle\frac{1}{8\pi}}\,P^2 \Big[-{\textstyle\frac{1}{2}}a_{,xx}+{\textstyle\frac{1}{2}}a_{,x}\Big(f-\frac{P_{,x}}{P}\Big)
-{\textstyle\frac{1}{4}}a\Big(f^2+\frac{4\Lambda}{P^2}\Big)
-{\textstyle\frac{1}{2}}b\Big(e_{,x}+\frac{P_{,x}}{P}\,e+ \frac{P_{,u}}{P^3}\Big) \nonumber\\
&&\hspace{2.0mm}
+fe\frac{P_{,u}}{P}+e_{,ux}+ \frac{P_{,x}}{P}\,e_{,u}
+(\Lambda+{\textstyle\frac{1}{4}}P^2f^2) e^2
+ \frac{P_{,uu}}{P^3}-2\frac{P_{,u}^2}{P^4}\Big] +2P^2 e\,\mathcal{J}\,. \label{NKundt2fnot=0sum}
\end{eqnarray}
The function $\mathcal{J}$ can be prescribed arbitrarily by choosing, e.g., (the spatial derivative of) any free function $b(u,x)$, while the function $\mathcal{N}$ can be prescribed by the free function $a(u,x)$.

Moreover, the general expressions (\ref{Jfnot=0sumar}) and (\ref{NKundt2fnot=0sum}) reduce to (\ref{Jf=0sumar}) and (\ref{Nf=0sumar}) when ${f=0}$ (in which case necessarily ${\Lambda=0}$).

The metric (\ref{Kundtmetric_fnot=0}) is a vacuum solution if, and only if, ${\mathcal{J}=0=\mathcal{N}}$. It is Minkowski, de~Sitter, or anti-de~Sitter space, according to the sign of the cosmological constant $\Lambda$.

\end{itemize}

In fact, as already identified in subsection 6.2 of~\cite{PodolskyZofka:2009} for general Kundt geometries in an arbitrary dimension, \emph{these are two distinct invariant subclasses} defined by the property ${f=0}$ and ${f \not=0}$ since the scalar
\begin{equation}
f^if_i\equiv g^{ij}f_if_j=g^{xx}f_xf_x=P^2f^2=\F
\label{invariantF}
\end{equation}
is gauge invariant. Moreover, this distinction is related to the (non)vanishing of the Ricci rotation (NP) coefficient  ${\tau\equiv k_{a;b}\,m^al^b}$. The null triad is given by Eq.~(\ref{triadexpl}), for which ${\tau=\frac{1}{2}Pf}$. Therefore, for ${f\not=0}$, we obtain ${\tau\not=0}$ which means that the privileged null congruence generated by ${\mathbf{k}=\mathbf{\partial}_r}$ has its \emph{internal spatial rotation}. This effect is absent if, and only if, ${f=0}$. Such an invariant classification is analogous to the division into the class of {\it pp\,}-waves and the class of Kundt waves within the Kundt geometry in the standard ${D=4}$. (See section~18.3 of \cite{GriffithsPodolsky:2009}.)

\subsection{Gauge freedom}
\label{Kundtgauge}

Further simplification  is achieved by employing the gauge freedom.
As in section~\ref{gaugeRT}, we consider the coordinate transformations ${(u,x)\to (u,x')}$ defined by ${x=x(u,x')}$. Then, the Kundt metric (\ref{Kundtmetric_fnot=0}) is transformed to
\begin{eqnarray}
\dd s^2 \rovno \frac{1}{P^2}\biggl(\frac{\partial x}{\partial x'}\biggl)^2\dd x'^2+2\frac{\partial x}{\partial x'}\biggl[\frac{1}{P^2}\frac{\partial x}{\partial u}+\,(e+f\,r)\biggl]\,\dd u \dd x' \nonumber \\
&&-2\,\dd u\dd r  +\Bigg[a+b\,r +\big(\Lambda+{\textstyle\frac{1}{4}} \F \big)\,r^2+\frac{1}{P^2}\biggl(\frac{\partial x}{\partial u}\biggl)^2+2\,(e+f\,r)\frac{\partial x}{\partial u} \Bigg]\, \dd u^2\,.
\end{eqnarray}
Two natural possibilities are:

\begin{itemize}
\item \textbf{Setting ${e=0}$}

The transformation (\ref{gauge of e}), if we redefine the functions such that ${P/\frac{\partial x}{\partial x'}\to P}$, ${\frac{\partial x}{\partial x'}f\to f}$, ${a-P^2e^2\to a}$, and ${b-2P^2ef\to b}$, locally removes the function $e$ from the metric (\ref{Kundtmetric_fnot=0}), and then the resulting Kundt metric (without a prime on the new coordinate $x'$) is given by
\begin{eqnarray}
\dd s^2 \rovno \frac{1}{P^2}\, \dd x^2+2\,fr\,\dd u \dd x -2\,\dd u\dd r  +\Big[a+b\,r +\big(\Lambda+{\textstyle\frac{1}{4}} \F \big)\,r^2 \Big]\, \dd u^2 \,, \label{Kundtmetric_fnot=0e=0}
\end{eqnarray}
with
\begin{eqnarray}
\mathcal{J}(u,x) \rovno {\textstyle\frac{1}{16\pi} \Big[\, f_{,u}-b_{,x}
 -f\,(\ln P)_{,u}\,\Big]}\,, \label{Jfnot=0sumare=0}\\
\mathcal{N}(u,x) \rovno {\textstyle\frac{1}{8\pi}}\,P^2 \Big[\!\!-{\textstyle\frac{1}{2}}a_{,xx}+{\textstyle\frac{1}{2}}a_{,x}\Big(f-\frac{P_{,x}}{P}\Big) \nonumber \\
&&\qquad\ -{\textstyle\frac{1}{4}}a\Big(f^2+\frac{4\Lambda}{P^2}\Big)-{\textstyle\frac{1}{2}}b\frac{P_{,u}}{P^3}
+ \frac{P_{,uu}}{P^3}-2\frac{P_{,u}^2}{P^4}\Big]\,. \label{NKundt2fnot=0sume=0}
\end{eqnarray}

\item \textbf{Setting ${P=1}$}

Alternatively, the transformation (\ref{gauge of P}) with redefinition of the functions ${e, f, a, b}$ such that ${P^{-1}\frac{\partial x}{\partial u}+Pe\to e}$, ${Pf\to f}$, ${a+\frac{1}{P^2}(\frac{\partial x}{\partial u})^2+2e\frac{\partial x}{\partial u}\to a}$, and ${b+2f\frac{\partial x}{\partial u}\to b}$, puts the Kundt metric (\ref{Kundtmetric_fnot=0}) into the form with ${P=1}$, namely
\begin{eqnarray}
\dd s^2 \rovno \dd x^2+2\,(e+f\,r)\,\dd u \dd x -2\,\dd u\dd r  +\Big[a+b\,r +\big(\Lambda+{\textstyle\frac{1}{4}} f^2 \big)\,r^2 \Big]\, \dd u^2 \,, \label{Kundtmetric_fnot=0P=1}
\end{eqnarray}
for which the field equations (\ref{Jfnot=0sumar}) and (\ref{NKundt2fnot=0sum}) are
\begin{eqnarray}
\mathcal{J}(u,x) \rovno {\textstyle\frac{1}{16\pi} \Big[\, f_{,u}-b_{,x}
 -\frac{1}{2}(f^2+4\Lambda)e\,\Big]}\,, \label{Jfnot=0sumarp=1}\\
\mathcal{N}(u,x) \rovno {\textstyle\frac{1}{8\pi}} \Big[\!\!-{\textstyle\frac{1}{2}}a_{,xx}+{\textstyle\frac{1}{2}}a_{,x}\,f
+{\textstyle\frac{1}{4}} (e^2-a)(f^2+4\Lambda)
-{\textstyle\frac{1}{2}}b \, e_{,x}
+e_{,ux} \Big] +2 e\,\mathcal{J}\,. \label{NKundt2fnot=0sumP=1}
\end{eqnarray}

\end{itemize}

\subsection{Important subclasses}
\label{Kundtsubclasses}

Finally, let us mention three important subclasses of the general Kundt metric:

\begin{itemize}
\item \textbf{pp-waves}

These are defined geometrically as admitting a \emph{covariantly constant null vector} field ${\mathbf{k}}$ (and thus are sometimes denoted as CCNV spacetimes). Here we have ${\mathbf{k}=\mathbf{\partial}_r\,}$ and, since ${k_{\alpha;\beta}=\frac12g_{\alpha\beta,r}}$, it follows that all the functions in the metric (\ref{Kundtmetric_fnot=0}) must be independent of the coordinate $r$. Therefore, it takes the Brinkmann form
\begin{eqnarray}
\dd s^2 \rovno \frac{1}{P^2}\, \dd x^2+2\,e\,\dd u \dd x -2\,\dd u\dd r  + a\, \dd u^2 \,, \label{pp-metric}
\end{eqnarray}
which is Eq.~(\ref{Kundtmetric_f=0}) with ${b=0}$, and necessarily ${\Lambda=0}$. From Eq.~(\ref{Jf=0sumar}) it immediately follows that ${\mathcal{J}=0}$, i.e., there are \emph{no gyratonic pp-waves in Einstein's theory in three dimensions}. Only the standard pure radiation is allowed. Using the gauge ${e=0}$, corresponding to the metric (\ref{Kundtmetric_fnot=0e=0}), this is given by the following single remaining field equation:
\begin{equation}
\mathcal{N}(u,x) = {\textstyle\frac{1}{8\pi}}\, \Big[-{\textstyle\frac{1}{2}}(P^2 a_{,xx}+PP_{,x} a_{,x})
+ \frac{P_{,uu}}{P} -2\frac{P_{,u}^2}{P^2} \Big]\,. \label{ppe=0}
\end{equation}

\item \textbf{VSI spacetimes}

VSI spacetimes are Lorentzian geometries with \emph{vanishing scalar curvature invariants of all orders}.  A review and list of references can be found in~\cite{OrtaggioPravdaPravdova:2013}.
It was found that all such spacetimes belong to the Kundt class and can be written in 2+1 dimensions in the following canonical form:
\begin{equation}
 \dd s^2=\dd x^2+2\,(e+ f\,r )\,\dd x\dd u-2\,\dd u\,\dd r+(a + b\,r + c\,r^2)\,\dd u^2 \,,
 \label{metricVSI}
\end{equation}
where ${a, b, c, e,f}$ are functions of $u$ and $x$. This is a particular subcase of the metric (\ref{Kundtmetric_fnot=0}) in which
\begin{equation}
 P=1\,,\qquad  c\equiv\Lambda+{\textstyle\frac{1}{4}} \F=\Lambda+{\textstyle\frac{1}{4}} f^2 \,.
 \label{flatspace}
\end{equation}
Again, two subclasses can now be distinguished, namely the case ${f=0}$  and the case ${f \not=0}$.
(For more details, see \cite{ColeyFusterHervikPelavas:2006, PodolskyZofka:2009}.) The special subclass of VSI spacetimes with ${f=0}$ and  ${b=0=c}$ represent pp-waves (\ref{pp-metric}) with a trivial spatial metric part. In view of Eq.~(\ref{ppe=0}), applying the gauge ${e=0}$, it is a pure-radiation spacetime
\begin{equation}
 \dd s^2=\dd x^2-2\,\dd u\,\dd r+a(x,u)\,\dd u^2
 \label{metricVSIe=0=f}
\end{equation}
with
\begin{equation}
\mathcal{N}(u,x) = -{\textstyle\frac{1}{16\pi}}\,  a_{,xx}\,. \label{VSIppe=0}
\end{equation}

\item \textbf{CSI spacetimes}

Moreover, in \cite{ColeyHervikPelavas:2008}, Lorentzian manifolds in $D=3$ were studied for which all polynomial scalar invariants constructed from the Riemann tensor and its covariant derivatives are \emph{constant}. These CSI geometries are either locally homogeneous or locally Kundt. They thus also belong to the metrics investigated in this contribution. Using our notation, the Kundt CSI geometries of \cite{ColeyHervikPelavas:2008} correspond to the line element
\begin{equation}
 \dd s^2=\dd x^2+2\,(e+ f\,r )\,\dd x\dd u-2\,\dd u\,\dd r+\Big[a + b\,r + \big(\sigma+{\textstyle \frac{1}{4}f^2}\big)\,r^2\Big]\,\dd u^2 \,,
 \label{metricCSI}
\end{equation}
with the metric function $f(u,x)$ constrained by
\begin{equation}
f_{,x}+{\textstyle \frac{1}{2}}f^2=-s \,, \qquad (2\sigma-s)f=-2\alpha \,,
\end{equation}
where ${\alpha}$, ${\sigma}$, and ${s}$ are constants. Comparing this geometric construction with the explicit solution of Einstein's equations (\ref{Kundtmetric_fnot=0}) and (\ref{Fsummary}), in the gauge (\ref{Kundtmetric_fnot=0P=1}), we identify ${\sigma=\Lambda}$ and we find ${s=2\Lambda}$, that is ${\alpha=0}$. This brings us to the Segre types ${\{3\}}$, ${\{(21)\}}$, and ${\{(1, 11)\}}$ of \cite{ColeyHervikPelavas:2008}.

\end{itemize}

\newpage

\section{Summary and future prospects}

In the present paper, we have obtained the general solution of the Einstein equations in 2+1 dimensions in the presence of a cosmological constant for a pure radiation (null dust) or gyratons under a few weak assumptions.
Our result is summarized as follows:

\vspace{5mm}
\noindent \textbf{Theorem~5:} If the spacetime admits a geodesic null vector field, the general local solution of the 3-dimensional Einstein equations with an aligned pure radiation or gyratons in the presence of a cosmological constant consists of the Robinson--Trautman class (\ref{RTmetricsummary}) with the non-zero components of the energy-momentum tensor (\ref{RTTuu}) and (\ref{RTTup}) satisfying Eqs.~(\ref{RTE1}) and (\ref{RTE2}), and the Kundt class (\ref{Kundtmetric_fnot=0}) with the non-zero components of the energy-momentum tensor (\ref{KundtTuu}) and (\ref{KundtTup}) satisfying Eqs.~(\ref{Jfnot=0sumar}), (\ref{NKundt2fnot=0sum}), and (\ref{Fsummary}). There are two invariantly distinct subclasses, namely ${f=0}$ and ${f\ne0}$.
\vspace{5mm}

As stated above, the general solution in three dimensions is divided into the Robinson--Trautman class and the Kundt class.
A characteristic feature of the Robinson--Trautman class is the fact that the function $f(u,x)$ can be non-zero, in spite that it has to be vanishing in four and higher dimensions.
This surprising fact shows a richer structure of the 3-dimensional solution than its higher-dimensional counterparts.
The spacetime with vanishing $f$ would share the same causal structure and physical interpretation with the higher-dimensional counterparts.
Actually, in the case of ${f=0}$, we have constructed the 3-dimensional version of the Kinnersley photon rocket in the absence of gyratons.
We leave it for future investigations to clarify a physical interpretation of the solution with non-vanishing $f$.

This new functional degree of freedom $f(u,x)$ exists in the Kundt class as well.
It would be interesting to compare all these explicit Kundt-type solutions in Einstein's theory with the analogous Kundt solutions of topologically massive gravity (a 3-dimensional theory with a cosmological constant, extending the Einstein gravity to contain higher derivative terms), as given in \cite{ChowPopeSezgin:2009} and summarized in Chapters~18 and~19 of \cite{GarciaDiaz:2017}. Also, comparison to higher-dimensional Kundt spacetimes in the Einstein theory in any dimension ${D\ge4}$,  investigated in \cite{PodolskyZofka:2009}, could reveal specific properties of these 2+1 spacetimes.

Our general solution could be used to provide answers to open problems in classical general relativity in the framework of 2+1 dimensions. An example of such problems is the cosmic censorship hypothesis~\cite{penrose1969,penrose1979} in the presence of a negative cosmological constant. In order to answer this fundamental question, we need to clarify whether the formation of a naked singularity is generic or not within our solution which describes gravitational collapse of a pure radiation or gyratons from regular initial data.

Another possible application of our solution is for canonical quantum gravity.
As shown in the ADM formalism of the Einstein equations, general relativity can be treated as a constrained dynamical system. There are two possible ways of canonical quantization for such systems: the Dirac quantization and the reduced phase-space quantization. (See~\cite{carlip} for a textbook of quantum gravity in 2+1 dimensions.) In the latter approach, one first solves the classical constraint equations, namely the Hamiltonian constraint and momentum constraint. Then, we put the solutions of these constraint equations back to the action, perform the spatial integration, and finally quantize the system on the constraint surface.
This is a natural way to quantize constrained dynamical systems, however, it is an extremely difficult task to solve the classical constraint equations with full generality.

In the present paper, we have succeeded to derive the general solution of the Einstein equations thanks to the very convenient canonical coordinate system adopted.
Such an appropriate coordinate system could give us a hint to solve constraint equations and perform a complete reduced phase-space quantization of 3-dimensional gravity.
If this is possible in the presence of a negative cosmological constant, in particular, the result would bring us a full description of the quantum states of 3-dimensional black holes.

\section*{Acknowledgments}
H.~M.~thanks the Institute of Theoretical Physics at the Charles University for a kind hospitality in Prague where this work was initiated, and J.~P.~thanks Hokkai-Gakuen University for a kind hospitality in Sapporo, where this work was completed.
This article was supported by the Czech Science Foundation grant No.~GA\v{C}R 17-01625S.

\appendix

\section{Connections and curvature components in canonical coordinates}
\label{sec_geomcurv}
\renewcommand{\theequation}{A\arabic{equation}}
\setcounter{equation}{0}

The Christoffel symbols for the general non-twisting spacetime (\ref{general nontwist}) after applying the condition (\ref{shearfree condition}) are
\begin{eqnarray}
&& {\textstyle \Gamma^r_{rr} = 0} \,, \label{ChristoffelBegin} \\
&& {\textstyle \Gamma^r_{ru} = -\frac{1}{2}g_{uu,r}+\frac{1}{2}g^{rx}g_{ux,r}} \,, \\
&& {\textstyle \Gamma^r_{rx} = -\frac{1}{2}g_{ux,r}+\Theta g_{ux}} \,, \\
&& {\textstyle \Gamma^r_{uu} = \frac{1}{2}\big[-g^{rr}g_{uu,r}-g_{uu,u}+g^{rx}(2g_{ux,u}-g_{uu,x})\big]} \,, \\
&& {\textstyle \Gamma^r_{ux} = \frac{1}{2}\big[-g^{rr}g_{ux,r}-g_{uu,x}+g^{rx}g_{xx,u}\big]} \,, \\
&& {\textstyle \Gamma^r_{xx} = -\Theta g^{rr}g_{xx}-g_{ux||x}+\frac{1}{2}g_{xx,u}} \,, \\
&& {\textstyle \Gamma^u_{rr}=\Gamma^u_{ru}=\Gamma^u_{rx} = 0} \,, \\
&& {\textstyle \Gamma^u_{uu} = \frac{1}{2}g_{uu,r}} \,, \\
&& {\textstyle \Gamma^u_{ux} = \frac{1}{2}g_{ux,r}} \,, \\
&& {\textstyle \Gamma^u_{xx} = \Theta g_{xx}} \,, \\
&& {\textstyle \Gamma^x_{rr} = 0} \,, \\
&& {\textstyle \Gamma^x_{ru} = \frac{1}{2}g^{xx}g_{ux,r}} \,, \\
&& {\textstyle \Gamma^x_{rx} = \Theta} \,, \\
&& {\textstyle \Gamma^x_{uu} = \frac{1}{2}\big[-g^{rx}g_{uu,r}+g^{xx}(2g_{ux,u}-g_{uu,x})\big]} \,, \\
&& {\textstyle \Gamma^x_{ux} = \frac{1}{2}\big[-g^{rx}g_{ux,r}+g^{xx}g_{xx,u}\big]} \,, \\
&& {\textstyle \Gamma^x_{xx} = -\Theta g^{rx}g_{xx}+\,^{S}\Gamma^x_{xx}} \,, \label{ChristoffelEnd}
\end{eqnarray}
where ${\,^{S}\Gamma^x_{xx}\equiv\frac{1}{2}g^{xx}g_{xx,x}}$ is the Christoffel symbol with respect to the only spatial coordinate $x$, i.e.,  coefficient of the covariant derivative on the transverse 1-dimensional space spanned by~$x$.

The non-vanishing Riemann curvature tensor components are then
\begin{eqnarray}
&& R_{rxrx} = {\textstyle -\big(\Theta_{,r}+\Theta^2\big)g_{xx}} \,, \\
&& R_{rxru} = {\textstyle -\frac{1}{2}g_{ux,rr}+\frac{1}{2}\Theta g_{ux,r}} \,, \\
&& R_{ruru} = {\textstyle -\frac{1}{2}g_{uu,rr}+\frac{1}{4}g^{xx}(g_{ux,r})^2} \,, \\
&& R_{rxux} = {\textstyle \frac{1}{2}g_{ux,r||x}+\frac{1}{4}(g_{ux,r})^2-g_{xx}\Theta_{,u}
-\frac{1}{2}\Theta\big(g_{xx,u}+g_{xx}g_{uu,r}\big)} \,, \\
&& R_{ruux} = {\textstyle g_{u[u,x],r}+\frac{1}{4}g^{rx}(g_{ux,r})^2-\frac{1}{4}g^{xx}g_{xx,u}g_{ux,r}} \nonumber \\
&& \hspace{15.0mm} {\textstyle +\Theta\big(g_{ux,u}-\frac{1}{2}g_{uu,x}-\frac{1}{2}g_{ux}g_{uu,r}\big)} \,, \\
&& R_{uxux} = {\textstyle -\frac{1}{2}(g_{uu})_{||xx}+g_{ux,u||x}-\frac{1}{2}g_{xx,uu}+\frac{1}{4}g^{rr}(g_{ux,r})^2} \nonumber \\
&& \hspace{15.0mm} {\textstyle -\frac{1}{2}g_{uu,r}e_{xx}+\frac{1}{2}g_{uu,x}g_{ux,r}-\frac{1}{2}g^{rx}g_{xx,u}g_{ux,r}+\frac{1}{4}g^{xx}(g_{xx,u})^2} \nonumber \\
&& \hspace{15.0mm} {\textstyle -\frac{1}{2}\Theta g_{xx}\big[g^{rr}g_{uu,r}+g_{uu,u}-g^{rx}(2g_{ux,u}-g_{uu,x})\big]} \,.
\end{eqnarray}
Finally, the components of the Ricci tensor are
\begin{eqnarray}
&& R_{rr} = {\textstyle -\big(\Theta_{,r}+\Theta^2\big)} \,, \label{Ricci rr}\\
&& R_{rx} = {\textstyle -\frac{1}{2}g_{ux,rr}+\frac{1}{2}\Theta g_{ux,r}+\big(\Theta_{,r}+\Theta^2\big)g_{ux}} \,, \label{Ricci rp} \\
&& R_{ru} = {\textstyle -\frac{1}{2}g_{uu,rr}+\frac{1}{2}g^{rx}g_{ux,rr}+\frac{1}{2}g^{xx}\big(g_{ux,r||x}+(g_{ux,r})^2\big)} \nonumber \\
&& \hspace{12.0mm} {\textstyle -\Theta_{,u}-\frac{1}{2}\Theta\big(g^{xx}g_{xx,u}+g^{rx}g_{ux,r}+g_{uu,r}\big)} \,, \label{Ricci ru} \\
&& R_{xx} = {\textstyle -g_{xx}\,g^{rr}\big(\Theta_{,r} +\Theta^2 \big) +2g_{xx}\big(\Theta_{,u}-g^{rx}\Theta_{,x}\big)+2g_{ux}\Theta_{,x}-f_{xx}} \nonumber \\
&& \hspace{12.0mm} {\textstyle +\Theta\big[2g_{ux||x}+2g_{ux,r}g_{ux}+g_{xx}\big(g_{uu,r}-2g^{rx}g_{ux,r}\big)
-2e_{xx}\big]} \,, \label{Ricci pq} \\
&& R_{ux} = {\textstyle -\frac{1}{2}g^{rr}g_{ux,rr}-\frac{1}{2}g_{uu,rx}+\frac{1}{2}g_{ux,ru}
-\frac{1}{2}g^{rx}\big[g_{ux,r||x}+(g_{ux,r})^2\big]} \nonumber \\
&& \hspace{12.0mm} {\textstyle +g^{xx}\big(\frac{1}{2}g_{ux,r}g_{ux||x}-\frac{1}{2}e_{xx}g_{ux,r}\big)+g_{ux}\Theta_{,u}}\nonumber \\
&& \hspace{12.0mm} {\textstyle +\Theta\big[g_{ux}g_{uu,r}-\frac{1}{2}(g_{uu}g_{ux,r}-g_{uu,x})-g_{ux,u}
+\frac{1}{2}g^{rx}g_{ux,r}g_{ux}+\frac{1}{2}g^{rx}g_{xx,u}\big]} \,, \label{Ricci up} \\
&& R_{uu} = {\textstyle -\frac{1}{2}g^{rr}g_{uu,rr}-g^{rx}g_{uu,rx}-\frac{1}{2}g^{xx}e_{xx}g_{uu,r}+g^{rx}g_{ux,ru}-\frac{1}{2}g^{xx}g_{xx,uu}} \nonumber \\
&& \hspace{12.0mm} {\textstyle +g^{xx}(g_{ux,u||x}-\frac{1}{2}g_{uu||xx})+\frac{1}{2}(g^{rr}g^{xx}-g^{rx}g^{rx})(g_{ux,r})^2+\frac{1}{2}g^{xx}g_{ux,r}g_{uu,x} +\frac{1}{4}(g^{xx} g_{xx,u})^2} \nonumber \\
&& \hspace{12.0mm} {\textstyle
+\frac{1}{2}\Theta\big[-g^{rx}(2g_{ux,u}-g_{uu,x}-g_{ux}g_{uu,r})+g_{uu}g_{uu,r}-g_{uu,u}\big]} \,, \label{Ricci uu}
\end{eqnarray}
and the Ricci scalar is
\begin{eqnarray}
&& R = {\textstyle g_{uu,rr}-2g^{rx}g_{ux,rr}-2g^{xx}g_{ux,r||x}-\frac{3}{2}g^{xx}(g_{ux,r})^2} \nonumber \\
&& \hspace{8.0mm} {\textstyle +2\Theta_{,r}\,g_{uu}+4\Theta_{,u} +2\Theta^2g_{uu} +\Theta(2g_{uu,r}+2g^{rx}g_{ux,r}+2g^{xx}g_{xx,u})} \,.
\end{eqnarray}
The symbol ${\,_{||}}$ denotes the covariant derivative with respect to $g_{xx}$\,:
\begin{eqnarray}
g_{ux||x} \rovno g_{ux,x}-g_{ux}\,^{S}\Gamma^{x}_{xx}  \,, \\
g_{ux,r||x} \rovno g_{ux,rx}-g_{ux,r}\,^{S}\Gamma^{x}_{xx}  \,, \\
g_{ux,u||x} \rovno g_{ux,ux}-g_{ux,u}\,^{S}\Gamma^{x}_{xx}  \,, \\
(g_{uu})_{||xx} \rovno g_{uu,xx}-g_{uu,x}\,^{S}\Gamma^{x}_{xx}\,,
\end{eqnarray}
where ${e_{xx}}$ and ${f_{xx}}$ are convenient shorthands defined as
\begin{eqnarray}
e_{xx} \!\!\!\!& \equiv &\!\!\!\! g_{u{x||x}}- {\textstyle \frac{1}{2}}g_{xx,u} \,, \label{exx}\\
f_{xx} \!\!\!\!& \equiv &\!\!\!\! g_{ux,r||x}+ {\textstyle \frac{1}{2}}(g_{ux,r})^2 \label{fxx}\,.
\end{eqnarray}
The expressions (\ref{Ricci rr})--(\ref{Ricci uu}) of the Ricci tensor enable us to write explicitly the gravitational field equations for any  ${D=3}$ Kundt or Robinson--Trautman spacetime.

\newpage



\section{Derivation of the Robinson--Trautman solutions}
\label{sec:RobinsonTrautman}
\renewcommand{\theequation}{B\arabic{equation}}
\setcounter{equation}{0}

Here we explicitly perform a step-by-step integration of the Einstein field equations (\ref{EinstinEq}) with a matter field given by Eqs.~(\ref{EqTup}) and (\ref{EqTuu}) in the case ${\Theta\neq 0}$.

\subsection{Integration of ${R_{rr}= 0}$}
From Eq.~(\ref{Ricci rr}) we get the explicit form of this equation
\begin{equation}
\Theta_{,r}+\Theta^2=0 \,, \label{feqrr}
\end{equation}
which obviously determines the $r$-dependence of the expansion scalar $\Theta$. Its general solution can be written as $\Theta^{-1}=r+r_{0}(u,x)$. However, the metric (\ref{general nontwist}) is invariant under the gauge transformation ${r \to r-r_{0}(u,x)}$ and we can thus, without loss of generality, set the integration function ${r_{0}(u,x)}$ to zero. The expansion simply becomes
\begin{equation}
\Theta=\frac{1}{r}\,. \label{ExplEx}
\end{equation}
The integral form (\ref{IntShearFreeCond}) of the condition (\ref{shearfree condition}) with the expansion given by Eq.~(\ref{ExplEx}) completely determines the $r$-dependence of the 1-dimensional spatial metric ${g_{xx}(r,u,x)}$, namely
\begin{equation}
g_{xx}=r^2\,h_{xx}(u,x) \,. \label{SpMetr}
\end{equation}
It is convenient to introduce a function $P(u,x)$ such that ${h_{xx}(u,x) \equiv P^{-2}(u,x)}$, so that the metric component (\ref{SpMetr}) takes the form of
\begin{equation}
g_{xx}=\frac{r^2}{P^2(u,x)} \,. \label{SpMetric}
\end{equation}
Of course, by inversion, ${h^{xx}=P^2}$ and ${g^{xx}=P^2\,r^{-2}}$.

\subsection{Integration of ${R_{rx}= 0}$}
Using Eqs.~(\ref{Ricci rp}) and (\ref{feqrr}), which implies Eq.~(\ref{ExplEx}), the Ricci tensor component ${R_{rx}}$ becomes
\begin{equation}
{\textstyle R_{rx}=-\frac{1}{2}\left(g_{ux,rr} - g_{ux,r}\,r^{-1} \right) } \,.
\label{feqrp}
\end{equation}
The field equation ${R_{rx}= 0}$ can thus be immediately integrated, yielding a general solution
\begin{equation}
g_{ux}= e(u,x)\,r^2+f(u,x) \,, \label{NediagCov}
\end{equation}
where $e$ and $f$ are arbitrary integration functions of $u$ and $x$.
In view of Eqs.~(\ref{CovariantMetricComp}) and (\ref{SpMetric}), the corresponding contravariant component of the Robinson--Trautman metric is
\begin{equation}
g^{rx}=P^2\big[e(u,x)+f(u,x)\,r^{-2} \big] \,. \label{NediagContra}
\end{equation}

Already at this stage we can also fully integrate the energy-momentum conservation equations (\ref{EqTup}) and (\ref{EqTuu}) which determine the $r$-dependence of the gyratonic-type energy-momentum tensor:
\begin{eqnarray}
&& T_{ux} = \mathcal{J}\,r^{-1} \,, \label{ExplTup} \\
&& T_{uu} = \mathcal{N}\,r^{-1}-P(P\mathcal{J})_{,x}\,r^{-2}+fP^2\mathcal{J}\,r^{-3} \,, \label{ExplTuu}
\end{eqnarray}
where ${\mathcal{J}(u,x)}$ and ${\mathcal{N}(u,x)}$ are integration functions of $u$ and $x$.

\subsection{Integration of ${R_{ru}= -2\Lambda}$}
Using Eqs.~(\ref{NediagCov}) and (\ref{NediagContra}), with Eqs.~(\ref{ExplEx}) and (\ref{SpMetric}), the next Ricci tensor component (\ref{Ricci ru}) becomes
\begin{eqnarray}
&& {\textstyle  R_{ru} =
 -\frac{1}{2}\big( r\, g_{uu,r}\big)_{,r}\,r^{-1}
 +P^2\big(e_{||x}-\frac{1}{2}h_{xx,u}\big)\,r^{-1} +2\,P^2e^2} \,. \label{Rru EEq}
\end{eqnarray}
Recall that
\begin{equation}
e_{||x} \equiv e_{,x}-{\textstyle\frac{1}{2}} g^{xx}g_{xx,x}\,e =
e_{,x} + (P_{,x}/P)\,e \,, \label{defeparder}
\end{equation}
from which we obtain useful identities
\begin{equation}
P\,e_{||x} = (Pe)_{,x} \,, \qquad
e\,P^2\,e_{||x} = {\textstyle \frac{1}{2}}(P^2e^2)_{,x} \,,  \label{idno2}
\end{equation}
and
\begin{equation}
 c \equiv 2P^2\big(e_{||x}-{\textstyle\frac{1}{2}} h_{xx,u}\big) =
2\big[P(Pe)_{,x}+(\ln P)_{,u}\big]\,.  \label{idno3}
\end{equation}
Using Eq.~(\ref{Rru EEq}), the corresponding Einstein equation ${R_{ru}= -2\Lambda}$ can be easily integrated to give
\begin{equation}
g_{uu}=-a-b\,\ln |r|+c\,r
+(\Lambda+P^2e^2)\,r^2 \,, \label{guuExpl}
\end{equation}
where $a(u,x)$ and $b(u,x)$ are arbitrary functions.
The $r$-dependence of all metric components is thus fully established.


\subsection{Integration of ${R_{xx}=2\Lambda\,g_{xx}}$}
Using Eqs.~(\ref{feqrr}), (\ref{ExplEx}), (\ref{SpMetric}), (\ref{NediagCov}) and (\ref{NediagContra}), the general Ricci tensor component (\ref{Ricci pq}) becomes
\begin{eqnarray}
&& {\textstyle R_{xx}= -2\big(e_{||x}-\frac{1}{2}h_{xx,u}\big)\,r -2\,e^2\,r^{2} + r\,P^{-2}\,g_{uu,r}} \,.
\end{eqnarray}
Substituting now the expression (\ref{guuExpl}), we obtain
${R_{xx} = 2\Lambda\,g_{xx}-b\,P^{-2}}$. The corresponding Einstein equation ${R_{xx}=2\Lambda\,g_{xx}}$ is thus satisfied if, and only if,
\begin{equation}
b=0 \,. \label{bje0}
\end{equation}

At this stage, the most general Robinson--Trautman solution takes the form
\begin{eqnarray}
\dd s^2 \rovno \frac{r^2}{P^2}\, \dd x^2+2\,(e\,r^2+f\,)\,\dd u \dd x -2\,\dd u\dd r \nonumber \\
&& +\Big(-a  +2\big[ P(Pe)_{,x}+(\ln P)_{,u} \big]\,r +(\Lambda+P^2e^2)\,r^2\Big)\, \dd u^2 \,. \label{RTmetric}
\end{eqnarray}
Interestingly, the metric function $f(u,x)$ \emph{remains arbitrary} and, in general, \emph{non-vanishing}.

\subsection{Integration of ${R_{ux}=2\Lambda\,g_{ux} +8\pi\,T_{ux}}$}
Using Eqs.~(\ref{ExplEx}), (\ref{SpMetric}), (\ref{NediagCov}), (\ref{NediagContra}), and (\ref{guuExpl}) with Eq.~(\ref{bje0}), the Ricci tensor component ${R_{ux}}$ given by Eq.~(\ref{Ricci up}) becomes
\begin{equation}
R_{ux}= 2\Lambda\,g_{ux}
+\Big[-{\textstyle \frac{1}{2}}a_{,x}-f_{,u}+f\,P^2\big(e_{||x}
-{\textstyle \frac{1}{2}} h_{xx,u}\big)\Big]\,r^{-1}\,.
\end{equation}
The field equation ${R_{ux}=2\Lambda\,g_{ux} +8\pi\,T_{ux}}$ with the gyratonic component (\ref{ExplTup}) of the energy-momentum tensor ${T_{ux} = \mathcal{J}\,r^{-1}}$ thus reduces to a simple equation
\begin{equation}
 a_{,x}+2f_{,u}-c\,f = -16\pi\, \mathcal{J} \,.
 \label{fieleq_ux}
\end{equation}
Clearly, this establishes a unique relation between the gyratonic matter source, represented by its angular momentum function $\mathcal{J}$, and the metric functions. In particular, it can be used as an equation determining the function $a(u,x)$ in the metric (\ref{RTmetric}).

\subsection{Integration of ${R_{uu}=2\Lambda\,g_{uu} +8\pi\, T_{uu}}$}

This final equation determines the relation between the Robinson--Trautman geometry and the pure radiation matter field represented by the profile $\mathcal{N}(u,x)$ in Eq.~(\ref{ExplTuu}). Using Eqs.~(\ref{ExplEx}), (\ref{SpMetric}), (\ref{NediagCov}), (\ref{NediagContra}), and (\ref{guuExpl}) with Eqs.~(\ref{bje0}) and (\ref{idno3}), the Ricci tensor component ${R_{uu}}$ given by Eq.~(\ref{Ricci uu}) becomes
\begin{eqnarray}
&& R_{uu}= \textstyle{ 2\Lambda\,g_{uu}
-\frac{1}{2} fP^2\big(a_{,x}+2 f_{,u}-cf \big)\,r^{-3}
+\frac{1}{2} P \big[P(a_{,x}+2 f_{,u}-cf)\big]_{,x} \,r^{-2}} \nonumber\\
&& \hspace{11.0mm} {\textstyle +\frac{1}{2}\,\big[\,a_{,u}-ac-P(Pc_{,x})_{,x}-2(\Lambda+P^2e^2)P(Pf)_{,x}-3P^2f(P^2e^2)_{,x} } \nonumber\\
&& \hspace{51.0mm} {\textstyle +2P^2f\,e_{,u}+P^2e\big(2cf -2 f_{,u}-3a_{,x}\big)\big]\,r^{-1}} \,,
\label{Ruu1}
\end{eqnarray}
where $c$ is given by Eq.~(\ref{idno3}). Applying the previous field equation (\ref{fieleq_ux}), the terms proportional to ${r^{-3}}$ and ${r^{-2}}$ are simply re-expressed using the gyratonic matter function $\mathcal{J}$. The field equation ${R_{uu}=2\Lambda\,g_{uu} +8\pi\, T_{uu}}$ with $T_{uu}$ given by Eq.~(\ref{ExplTuu}) then takes the form
\begin{eqnarray}
&&
 8\pi\, fP^2\mathcal{J}\,r^{-3}
-8\pi\, P(P\mathcal{J})_{,x} \,r^{-2} \nonumber\\
&& \hspace{3.0mm} {\textstyle +\frac{1}{2}}\,\big[\,a_{,u}-ac-P(Pc_{,x})_{,x}-2(\Lambda+P^2e^2)P(Pf)_{,x}  \nonumber\\
&& \hspace{10.0mm} -3P^2f(P^2e^2)_{,x}+2P^2f\,e_{,u}+P^2e\big(4f_{,u}-c f+48\pi \mathcal{J}\big)\big]\,r^{-1} \nonumber\\
&& =
 8\pi\, fP^2\mathcal{J}\,r^{-3}
-8\pi\, P(P\mathcal{J})_{,x} \,r^{-2}
+8\pi\, \mathcal{N}\,r^{-1}\,.
\label{Ruu1subs}
\end{eqnarray}
Remarkably, the terms  proportional to ${r^{-3}}$ and ${r^{-2}}$  on both sides of this equation are \emph{the same}, so that we  obtain only one additional condition determined by the remaining terms proportional to ${r^{-1}}$, namely
\begin{eqnarray}
 a_{,u}\rovno ac+\triangle c+2(\Lambda+P^2e^2)P(Pf)_{,x} +3P^2f(P^2e^2)_{,x}  \nonumber\\
&& -2P^2f\,e_{,u} -P^2e(4 f_{,u}-cf+48\pi\, \mathcal{J})
+ 16\pi\,\mathcal{N} \,, \label{RTEq2}
\end{eqnarray}
where
\begin{equation}
\triangle c \equiv h^{xx}\,c_{||xx} = P(Pc_{,x})_{,x}
\label{Laplace}
\end{equation}
is the covariant Laplace operator on the 1-dimensional transverse Riemannian space spanned by~$x$, applied on the function~$c$.

\section{Derivation of the Kundt solutions}
\label{sec:Lundt}
\renewcommand{\theequation}{C\arabic{equation}}
\setcounter{equation}{0}

Finally, we explicitly integrate the Einstein equations (\ref{EinstinEq}) with the matter field given by Eqs.~(\ref{EqTup}) and (\ref{EqTuu}) in the case ${\Theta=0}$.

\subsection{Integration of ${R_{rr}= 0}$}
In view of Eq.~(\ref{Ricci rr}), this equation is trivially satisfied, and the 1-dimensional spatial metric ${g_{xx}}$ is $r$-independent:
\begin{equation}
g_{xx}=h_{xx}(u,x)\equiv P^{-2}(u,x) \,. \label{KSpMetr}
\end{equation}
Of course, ${g^{xx}=h^{xx}=P^2}$ holds.

\subsection{Integration of ${R_{rx}= 0}$}
The Ricci tensor component (\ref{Ricci rp}) reduces to ${R_{rx}=-\frac{1}{2}g_{ux,rr}}$. We thus obtain a general solution
\begin{equation}
g_{ux}= e(u,x) + f(u,x)\,r \,, \label{KNediagCov}
\end{equation}
where $e$ and $f$ are arbitrary functions of $u$ and $x$.
In view of Eqs.~(\ref{CovariantMetricComp}) and (\ref{KSpMetr}), the corresponding contravariant component of the Kundt metric is
\begin{equation}
g^{rx}=P^2\big[ e(u,x) + f(u,x)\,r \big] \,. \label{KNediagContra}
\end{equation}

The energy-momentum conservation equations (\ref{EqTup}) and (\ref{EqTuu}) can now be integrated to give
\begin{eqnarray}
&& T_{ux} = \mathcal{J} \,, \label{KExplTup} \\
&& {\textstyle T_{uu} = P^2\big(\mathcal{J}_{||x}+ f\mathcal{J}\big)\,r+\mathcal{N}} \,, \label{KExplTuu}
\end{eqnarray}
where ${\mathcal{J}(u,x)}$ and ${\mathcal{N}(u,x)}$ are any functions of $u$ and $x$, and
${\mathcal{J}_{||x} \equiv \mathcal{J}_{,x} + (P_{,x}/P)\,\mathcal{J}}$.

\subsection{Integration of ${R_{ru}= -2\Lambda}$}

Using Eqs.~(\ref{KSpMetr}) and (\ref{KNediagCov}), the Ricci tensor component (\ref{Ricci ru}) is
${R_{ru} = -\frac{1}{2}\, g_{uu,rr} +\frac{1}{2} P^2(f_{||x}+f^2)}$, where
\begin{equation}
f_{||x} \equiv f_{,x} + \frac{P_{,x}}{P}\,f
\quad\Leftrightarrow\quad Pf_{||x} \equiv (Pf)_{,x}\,.
\label{f||x}
\end{equation}
The corresponding Einstein equation gives
\begin{equation}
g_{uu}= a(u,x)+b(u,x)\,r +\big[2\Lambda+{\textstyle\frac{1}{2}} P^2(f_{||x}+f^2)\big]\,r^2 \,, \label{KguuExpl}
\end{equation}
where $a(u,x)$ and $b(u,x)$ are arbitrary functions.

\subsection{Integration of ${R_{xx}=2\Lambda\,g_{xx}}$}

For ${\Theta=0}$, using Eqs.~(\ref{fxx}) and (\ref{KNediagCov}), the Ricci tensor component (\ref{Ricci pq}) reduces to $R_{xx}= -f_{xx}=-\big(f_{||x}+ \frac{1}{2}f^2\big)$.
The field equation ${R_{xx}=2\Lambda\,g_{xx}}$ takes the form ${P^2(f_{||x}+ {\textstyle\frac{1}{2}}f^2)=-2\Lambda}$, i.e., using Eq.~(\ref{f||x}),
\begin{equation}
P^2f_{,x} + PP_{,x}\,f + {\textstyle\frac{1}{2}}P^2f^2=-2\Lambda \,. \label{fLambdaP}
\end{equation}
From this we can express $f_{||x}$ and substitute it into Eq.~(\ref{KguuExpl}), obtaining an explicit coefficient ${g_{uu}}$. At this stage, the most general Kundt solution in ${D=3}$ becomes
\begin{eqnarray}
\dd s^2 \rovno \frac{1}{P^2}\, \dd x^2+2\,(e+f\,r)\,\dd u \dd x -2\,\dd u\dd r  +\Big[a+b\,r +\big(\Lambda+{\textstyle\frac{1}{4}} \F \big)\,r^2 \Big]\, \dd u^2 \,, \label{Kundtmetric}
\end{eqnarray}
where
\begin{equation}
\F \equiv P^2f^2\,.
\label{defF}
\end{equation}
The functions $f$ and $P$ are coupled by the field equation (\ref{fLambdaP}), which can be integrated as follows:

\begin{itemize}
\item \textbf{The case} ${f=0\Leftrightarrow F=0}$: Necessarily ${\Lambda=0}$, while ${P\not=0}$ remains arbitrary.

\item \textbf{The case} ${f\not=0\Leftrightarrow F\not=0}$: The field equation (\ref{fLambdaP}) can be rewritten using Eq.~(\ref{defF}) as
\begin{equation}
 \F_{,x}+ (\F+4\Lambda)f=0 \,. \label{F}
\end{equation}
Either we can treat $F(u,x)$ as an \emph{arbitrary generating function}, from which we obtain ${f=-\F_{,x}/(\F+4\Lambda)}$ and then ${P^2=\F/f^2=\F(\F+4\Lambda)^2/(\F_{,x})^2}$, or we prescribe any $f(x,u)$ and integrate Eq.~(\ref{F}) to ${\F=-4\Lambda+g(u)\,\exp[-\int\! f(x,u)\,\dd x]}$, where $g(u)$ is an arbitrary function of $u$, and then ${P^2=\F/f^2}$. In both cases, the field equation (\ref{fLambdaP}) is satisfied.
\end{itemize}

\subsection{Integration of ${R_{ux}=2\Lambda\,g_{ux} +8\pi\,T_{ux}}$}
Using Eq.~(\ref{Ricci up}) with ${\Theta=0}$ for the metric (\ref{Kundtmetric}), and applying the field equation  (\ref{fLambdaP}), we obtain
${ R_{ux} = \frac{1}{2}\big(\, f_{,u} -f\,(\ln P)_{,u}-b_{,x}+\frac{1}{2}(4\Lambda-\F)e\,\big)
-\frac{1}{4}\big(\F_{,x} + (\F-4\Lambda)f\,\big)\,r}$. Using Eqs.~(\ref{KNediagCov}) and (\ref{KExplTup}), the corresponding equation ${R_{ux}=2\Lambda\,g_{ux} +8\pi\,T_{ux}=2\Lambda\,(e+f\,r) +8\pi\,\mathcal{J}}$ splits into two equations, namely the coefficients for the powers $r^1$ and $r^0$.
The former equation is always satisfied due to Eq.~(\ref{F}), while the latter one is
\begin{equation}
\mathcal{J} = {\textstyle\frac{1}{16\pi} \big(\, f_{,u}-b_{,x}
 -\frac{1}{2}(F+4\Lambda)e-f\,(\ln P)_{,u}\,\big)} \,. \label{Rux2}
\end{equation}
It explicitly determines the gyratonic matter function $\mathcal{J}$ in terms of the metric functions representing the gravitational field.

\subsection{Integration of ${R_{uu}=2\Lambda\,g_{uu} +8\pi\,T_{uu}}$}
It remains to satisfy the last equation for the Ricci tensor component (\ref{Ricci uu}). Setting ${\Theta=0}$ and using the Kundt metric functions (\ref{Kundtmetric}) with Eq.~(\ref{F}), we obtain
\begin{equation}
R_{uu}= {\textstyle A+(\Lambda-\frac{1}{4}\F)\,a+\big(B+2\Lambda \,b\big)\,r
   +2\Lambda\big(\Lambda+\frac{1}{4}\F \big)r^2}\,,
\label{RuuKundt}
\end{equation}
where
\begin{eqnarray}
A \rovno
P^2 \Big[
-{\textstyle\frac{1}{2}}a_{,xx}+{\textstyle\frac{1}{2}}a_{,x}\Big(f-\frac{P_{,x}}{P}\Big)
-{\textstyle\frac{1}{2}}b\Big(e_{,x}+\frac{P_{,x}}{P}\,e+ \frac{P_{,u}}{P^3}\Big)  \nonumber\\
&& \hspace{4.5mm}
+{\textstyle\Big(f_{,u}-b_{,x}-\frac{1}{4}(F+4\Lambda)\,e\Big)e}+\Big(e_{,ux}+ \frac{P_{,x}}{P}\,e_{,u}\Big)
+ \frac{P_{,uu}}{P^3}-2\frac{P_{,u}^2}{P^4}\Big]\,,  \label{A}\\
B \rovno
P^2 \Big[-{\textstyle\frac{1}{2}}b_{,xx}+{\textstyle\frac{1}{2}}b_{,x}\Big(f-\frac{P_{,x}}{P}\Big)
+\Big(f_{,ux}+ \frac{P_{,x}}{P}\,f_{,u}\Big)
 \nonumber\\
&& \hspace{4.5mm}
+\big(f_{,u}-b_{,x}\big)f-\big(\Lambda+{\textstyle\frac{1}{4}}\F \big)\Big(e_{,x}+\frac{P_{,x}}{P}\,e+ \frac{P_{,u}}{P^3}\Big)\Big] \,.
\label{B}
\end{eqnarray}
Applying Eq.~(\ref{KExplTuu}) in the field equation  $R_{uu}=2\Lambda\,g_{uu} +8\pi\,T_{uu}=2\Lambda\,a+2\Lambda\,b\,r +2\Lambda\big(\Lambda+\frac{1}{4} \F \big)r^2+8\pi \big[P^2(\mathcal{J}_{||x}+ f\mathcal{J})r+\mathcal{N}\big]$, the quadratic term $r^2$ vanishes identically, while the  powers $r^1$ and $r^0$ yield the following two equations
\begin{eqnarray}
B \rovno 8\pi P^2\big(\mathcal{J}_{||x}+ f\mathcal{J}\big)\,,  \label{RuuKundt1}\\
A -\big(\Lambda+{\textstyle\frac{1}{4}}\F \big)a \rovno 8\pi\, \mathcal{N}\,. \label{RuuKundt2}
\end{eqnarray}
Evaluating the gyratonic term on the right hand side of Eq.~(\ref{RuuKundt1}) using Eqs.~(\ref{Rux2}) and (\ref{F}), we obtain
\begin{eqnarray}
8\pi P^2\big(\mathcal{J}_{||x}+ f\mathcal{J}\big) \rovno
P^2 \Big[ {\textstyle\frac{1}{2}}f_{,ux}-{\textstyle\frac{1}{2}}b_{,xx}
+\big(\Lambda+{\textstyle\frac{1}{4}}\F \big)\big(ef-e_{,x}\big)-{\textstyle\frac{1}{2}}f_{,x}\frac{P_{,u}}{P}
-{\textstyle\frac{1}{2}}f\Big(\frac{P_{,u}}{P}\Big)_{,x}  \nonumber\\
&&\hspace{4.0mm}
+\Big({\textstyle\frac{1}{2}}(f_{,u}-b_{,x})-(\Lambda+{\textstyle\frac{1}{4}}F)e-{\textstyle\frac{1}{2}}f\,\frac{P_{,u}}{P}\Big)
\Big(f+\frac{P_{,x}}{P}\Big)\Big] \,. \label{gyratmatterterm}
\end{eqnarray}
Interestingly, a straightforward but somewhat lengthy calculation using Eq.~(\ref{gyratmatterterm}), Eq.~(\ref{F}) and its $u$-derivative, shows that  the field equation (\ref{RuuKundt1}) is \emph{identically satisfied}.
It only remains to satisfy the last equation (\ref{RuuKundt2}).
We discuss two invariantly distinct subclasses separately:

\begin{itemize}
\item \textbf{The case} ${f=0}$: We have already proved that this necessarily requires ${\Lambda=0}$ and ${\F=0}$, and the gyratonic matter (\ref{Rux2}) is thus simply given by
\begin{equation}
\mathcal{J} = -{\textstyle\frac{1}{16\pi}}\,b_{,x}\,,
\label{JKundtf=0}
\end{equation}
while the  remaining equation (\ref{RuuKundt2}) reduces to
\begin{eqnarray}
&&
\mathcal{N}= {\textstyle\frac{1}{8\pi}}\,P^2 \Big[-{\textstyle\frac{1}{2}}\Big(a_{,xx}+a_{,x}\frac{P_{,x}}{P}\Big)
 -{\textstyle\frac{1}{2}}b\Big(e_{,x}+\frac{P_{,x}}{P}\,e+ \frac{P_{,u}}{P^3}\Big) \nonumber\\
&&\hspace{34.0mm}
+e_{,ux}+ \frac{P_{,x}}{P}\,e_{,u}
+ \frac{P_{,uu}}{P^3}-2\frac{P_{,u}^2}{P^4}\Big]+2P^2 e\,\mathcal{J} \,. \label{RuuKundt2f=0}
\end{eqnarray}
It explicitly gives the pure radiation profile function $\mathcal{N}$ in terms of the metric functions of the gravitational field. While $\mathcal{J}$ is given just by $b_{,x}$, the matter function $\mathcal{N}$ is given primarily by the spatial derivatives of $a$.

\item \textbf{The case} ${f\not=0}$:
The gyratonic matter ${T_{ux} = \mathcal{J} }$ is given by Eq.~(\ref{Rux2}), and the equation (\ref{RuuKundt2}) determines the pure radiation component $T_{uu}$ given by Eq.~(\ref{KExplTuu}). The corresponding profile function $\mathcal{N}$ is obtained from Eq.~(\ref{RuuKundt2}) using Eqs.~(\ref{A}) and (\ref{Rux2}) as
\begin{eqnarray}
\mathcal{N} \rovno {\textstyle\frac{1}{8\pi}}\,P^2 \Big[-{\textstyle\frac{1}{2}}a_{,xx}+{\textstyle\frac{1}{2}}a_{,x}\Big(f-\frac{P_{,x}}{P}\Big)
-{\textstyle\frac{1}{4}}a\Big(f^2+\frac{4\Lambda}{P^2}\Big)
-{\textstyle\frac{1}{2}}b\Big(e_{,x}+\frac{P_{,x}}{P}\,e+ \frac{P_{,u}}{P^3}\Big) \nonumber\\
&&\hspace{4.0mm}
+fe\frac{P_{,u}}{P}+e_{,ux}+ \frac{P_{,x}}{P}\,e_{,u}
+(\Lambda+{\textstyle\frac{1}{4}}P^2f^2) e^2
+ \frac{P_{,uu}}{P^3}-2\frac{P_{,u}^2}{P^4}\Big]+2P^2 e\,\mathcal{J} \,. \label{RuuKundt2fnot=0}
\end{eqnarray}
Such pure radiation can be prescribed arbitrarily by choosing any free function $a(u,x)$. Clearly, the expression (\ref{RuuKundt2fnot=0}) reduces to Eq.~(\ref{RuuKundt2f=0}) when ${f=0}$, implying ${\Lambda=0}$.
\end{itemize}

\end{document}